\documentclass[pmlr]{jmlr}

\RequirePackage{graphicx}
 \usepackage{booktabs}
\usepackage{longtable}
\makeatletter
\def\set@curr@file#1{\def\@curr@file{#1}} 
\makeatother

\theorembodyfont{\upshape}
\theoremheaderfont{\scshape}
\theorempostheader{:}
\theoremsep{\newline}

\jmlrvolume{298}
\jmlryear{2025}
\jmlrworkshop{Machine Learning for Healthcare}

\usepackage{multirow}
\usepackage{threeparttable}
\usepackage{makecell}
\usepackage[super]{nth}

\usepackage{siunitx}
\DeclareSIUnit\rbe{GyRBE}

\title[Patient-specific DRL for Automatic Replanning in HNC Proton Therapy]{Patient-Specific Deep Reinforcement Learning for Automatic Replanning in Head-and-Neck Cancer Proton Therapy}

\author{
       \Name{Malvern Madondo}\\
       \addr Department of Radiation and Cellular Oncology\\
       University of Chicago\\
       Chicago, IL, USA 
       \AND
       \Name{Yuan Shao}\\ 
       \addr Division of Environmental and Occupational Health Sciences\\
       University of Illinois at Chicago\\
       Chicago, IL, USA 
       \AND
       \Name{Yingzi Liu}\\ 
       \addr Department of Radiation and Cellular Oncology\\
       University of Chicago\\
       Chicago, IL, USA    
       \AND
       \Name{Jun Zhou}\\ 
       \addr Department of Radiation Oncology and Winship Cancer Institute\\
       Emory University\\
       Atlanta, GA, USA
       \AND
       \Name{Xiaofeng Yang}\\ 
       \addr Department of Radiation Oncology and Winship Cancer Institute\\
       Emory University\\
       Atlanta, GA, USA
       \AND
       \Name{Zhen Tian}
       \thanks{Corresponding author: Zhen Tian, PhD; email: ztian@bsd.uchicago.edu}
       \Email{\href{mailto:ztian@bsd.uchicago.edu}{\color{black}{ztian@bsd.uchicago.edu}}}\\ 
       \addr Department of Radiation and Cellular Oncology\\
       University of Chicago\\
       Chicago, IL, USA 
    } 

\begin{document}

\maketitle

\begin{abstract}
  Anatomical changes in head-and-neck cancer (HNC) patients during intensity-modulated proton therapy (IMPT) can shift the Bragg Peak of proton beams, risking tumor underdosing and organ-at-risk (OAR) overdosing. As a result, treatment replanning is often required to maintain clinically acceptable treatment quality. However, current manual replanning processes are often resource intensive and time consuming. In this work, we propose a patient-specific deep reinforcement learning (DRL) framework for automated IMPT replanning, with a reward-shaping mechanism based on a $150$-point plan quality score designed to handle competing clinical objectives in radiotherapy planning. We formulate the planning process as a reinforcement learning (RL) problem where agents learn high-dimensional control policies to adjust plan optimization priorities to maximize plan quality. Unlike population-based approaches, our framework trains personalized agents for each patient using their planning Computed Tomography (CT) and augmented anatomies simulating anatomical changes (tumor progression and regression). This patient-specific approach leverages anatomical similarities along the treatment course, enabling effective plan adaptation. We implemented and compared two DRL algorithms, Deep Q-Network (DQN) and Proximal Policy Optimization (PPO), using dose-volume histograms (DVHs) as state representations and a $22$-dimensional action space of priority adjustments. Evaluation on eight HNC patients using actual replanning CT data showed that both DRL agents improved initial plan scores from $120.78 \pm 17.18$ to $139.59 \pm 5.50$ (DQN) and $141.50 \pm 4.69$ (PPO), surpassing the replans manually generated by a human planner ($136.32 \pm 4.79$). Further comparison of dosimetric endpoints confirms these improvements translate to better tumor coverage and OAR sparing across diverse anatomical changes. This work highlights the potential of DRL in addressing the geometric and dosimetric complexities of adaptive proton therapy, offering a promising solution for efficient offline adaptation and paving the way for online adaptive proton therapy.
\end{abstract}

\section{Introduction}
\label{sec:intro}

Intensity-modulated proton therapy (IMPT) provides highly conformal tumor coverage while sparing surrounding organs at risk (OARs), leveraging the unique dose deposition characteristics of the proton beam’s Bragg peak~\citep{holliday2015proton, mckeever2016reduced, moreno2019intensity}. However, the Bragg peak makes IMPT highly sensitive to interfractional anatomical changes, such as tumor progression or regression, weight loss, and edema, that alter tissue density along the beam path~\citep{huiskes2023dosimetric, sonke2019adaptive}. These anatomical changes can shift the Bragg peak, degrading tumor coverage and/or OAR sparing, and often necessitate one or multiple manual replanning sessions during the treatment course to maintain clinical quality. Addressing these anatomical changes currently requires manual intervention: a multi-step process involving patient reimaging, anatomical recontouring, and iterative, trial-and-error plan optimization. This manual adaptation workflow is notoriously labor intensive and typically takes several days, delaying optimal care and straining limited clinical resources~\citep{kim2018proton, leeman2017proton, li2020past}. 

From a machine learning (ML) perspective, automating treatment planning presents several challenges. First, the planning process involves navigating a high-dimensional optimization space with complex, non-linear relationships between planning parameters and resulting dose distributions~\citep{burlacu2025deep, wildman2024recent}. Second, the multi-objective nature of treatment planning demands careful balancing of target coverage with the sparing of multiple OARs, each governed by its clinical constraints, thereby desiring methods that can effectively navigate trade-offs along a complex Pareto frontier~\citep{barker2004quantification, bobic2023large, sonke2019adaptive}. Finally, the limited availability of patient data and substantial inter-patient variability in anatomy and tumor characteristics make it difficult to develop models that generalize well across large patient populations~\citep{nikou2024modelling, visak2024assessing, volpe2021machine}. These challenges are further amplified in head and neck cancer (HNC) treatment planning, which involves multiple treatment targets with varying prescription dose levels, and tumors often invade or abut several critical organs.

Reinforcement learning (RL) offers a compelling framework for tackling these sequential, multi-objective optimization challenges in HNC treatment planning. RL agents learn optimal strategies through interaction and reward feedback, making them particularly well-suited for navigating the complex, iterative process of balancing competing clinical objectives in treatment planning~\citep{sutton1998reinforcement}. Specifically, RL holds the potential to automate the iterative decision-making process of priority adjustments in treatment planning~\citep{eckardt2021reinforcement, ebrahimi2021reinforcement, moreau2021reinforcement, yang2024automated}. Furthermore, RL's inherent adaptability aligns with the vision of personalized online adaptive radiotherapy, where treatment plans can be dynamically adapted based on a patient's tumor response and anatomical changes along the treatment course.

To realize this potential for automated IMPT replanning, we develop a patient-specific deep reinforcement learning (DRL) framework for the dosimetrically challenging HNC. Our approach formulates the priority tuning process during plan optimization as an RL problem, where the state is defined by dose–volume histograms (DVHs) for clinical target volumes (CTVs) and OARs, and the action space comprises $22$ predefined, clinically informed priority adjustments, enabling the agent to navigate the trade-offs inherent in this multi-objective optimization problem. Plan quality is quantified via a comprehensive $150$-point scoring system that combines ProKnow standardized scoring criteria~\citep{nelms2012variation} with institutional planning guidelines, and the reward is defined as the change in the plan quality score. Our framework trains personalized DRL agents using each patient’s anatomy captured by the initial planning Computed Tomography (CT) images and augmented anatomies simulating anatomical variations. This patient-specific approach leverages the patient's inherent anatomical consistency along the treatment course to optimize performance specifically for that individual rather than attempting to develop a one-size-fits-all solution.

\subsection*{Generalizable Insights about Machine Learning in the Context of Healthcare}

Our work offers broader insights for applying machine learning in healthcare settings characterized by high variability and limited data:
\begin{itemize}
    \item \textbf{Patient-specific architectures merit exploration in high-variability clinical applications.} In medical settings characterized by limited data availability and substantial anatomical variability, developing ML models that generalize across large patient cohorts remains a fundamental challenge. However, despite undergoing anatomical changes, each patient typically maintains high anatomical consistency throughout the treatment course, which is a feature that has been entirely underutilized in traditional population-based ML models. By simulating plausible anatomical variations from a patient’s baseline anatomy to augment the limited patient-specific training datasets, our proposed patient-specific strategy offers a viable alternative to traditional population-based ML models and aligns closely with the goals of personalized precision medicine.
    
    \item \textbf{Clinically-informed reward design ensures alignment.} Directly incorporating established clinical planning guidelines and quantitative plan scoring metrics into the RL reward function is crucial. This ensures that learned policies optimize for objectives recognized as clinically valid and relevant to real-world clinical decision-making priorities, fostering trust and clinical utility. 
\end{itemize}

\begin{table}[t]
    \centering
    \caption{\textbf{Planning objectives of IMPT inverse plan optimization for HNC}, including dose-volume constraints for CTVs and OARs. Here, $V_{d}$ is the volume receiving at least dose $d$, and $D_{v}$ is the minimum dose received by the hottest $v$ volume of a structure. \textit{Note}:~\si{\rbe} = dose (Gy) $\times$ relative biological effectiveness (RBE, typically 1.1 for protons).}
    \label{tab:planning_constraints}
    \resizebox{0.8\textwidth}{!}{%
        \begin{tabular}{ll}
            \toprule
            \textbf{Structure} & \textbf{Planning Objective} \\
            \midrule
            CTV1 (primary CTV) & 
              $\begin{aligned}
                & V_{d_{Rx,\text{CTV}_1}} \geq 98\%\ \text{of CTV1 volume} \\
                & D_{0\%} \leq 110\%\ \text{of}\ d_{Rx,\text{CTV}_1}
              \end{aligned}$ \\
            CTV2 (secondary CTV) & $V_{d_{Rx,\text{CTV}_2}} \geq 98\%\ \text{of CTV2 volume}$ \\
            CTV3 (tertiary CTV) & $V_{d_{Rx,\text{CTV}_3}} \geq 98\%\ \text{of CTV3 volume}$ \\
            Brainstem (BRS) & $D_{0.03\si{cc}} \leq 30~\si{\rbe}$ \\
            Spinal Cord (SC) & $D_{0.03\si{cc}} \leq 30~\si{\rbe}$ \\
            Mandible (MAN) & $V_{70\si{\rbe}} \leq 10\%\ \text{of MAN volume}$ \\
            Larynx (LAR) & $D_{\text{mean}}\leq 45~\si{\rbe}$ \\
            Pharynx (PHY) & $D_{\text{mean}} \leq 50~\si{\rbe}$ \\
            Left Parotid (PARL) & $D_{\text{mean}} \leq 26~\si{\rbe}$ \\
            Right Parotid (PARR) & $D_{\text{mean}} \leq 26~\si{\rbe}$ \\
            Left Cochlea (COCHL) & $D_{\text{mean}} \leq 35~\si{\rbe}$ \\
            Right Cochlea (COCHR) & $D_{\text{mean}} \leq 35~\si{\rbe}$ \\
            Left Submandibular Gland (SMGL) & $D_{\text{mean}} \leq 35~\si{\rbe}$ \\
            Right Submandibular Gland (SMGR) & $D_{\text{mean}} \leq 35~\si{\rbe}$ \\
            Esophagus (ESO) & $D_{\text{mean}} \leq 40~\si{\rbe}$ \\
            \bottomrule
        \end{tabular}
    }
\end{table}

\section{Related Work}
\label{sec:related}

RL has increasingly shown promise for automating radiotherapy treatment planning. Early demonstrations in cervical cancer high-dose-rate brachytherapy introduced a DRL-based Virtual Treatment Planner (VTP) that emulated human planners by observing DVH inputs and adjusting parameter weights, yielding plans that outperformed both the initial and human-generated plans~\citep{shen2019intelligent}. Building on this success, the framework was extended to external beam radiotherapy for prostate cancer~\citep{shen2020operating}, where separate DQN subnetworks adapted parameters in a relatively simple scenario with a single target and two OARs. Subsequent efforts by the same research group focused on improving training efficiency through rule-based adjustments informed by human-planner experience~\citep{shen2021improving} and scalability through a hierarchical VTP network that decomposed the planning process into structure selection, parameter selection, and action adjustment~\citep{shen2021hierarchical}, although still limited to single-target cases with few OARs. Diverging from neural network-based approaches, \citet{Zhang2020AnIP} developed an interpretable planning bot for pancreatic stereotactic body radiation therapy using linear function approximation for Q-learning. However, these approaches were primarily designed for anatomically simpler treatment sites and might struggle with sites like HNC, characterized by complex inter-structure trade-off relationships.

HNC represents a substantially more challenging planning context due to multiple target volumes with varying prescription doses and numerous critical structures in close proximity, demanding tight dosimetric trade-offs even for experienced planners. Recent RL adaptations to HNC planning have attempted to navigate this high-dimensional optimization space with varied strategies. \citet{gao2024human} adapted the hierarchical VTP approach for HNC \textit{photon} therapy (whereas our work focuses on \textit{proton} therapy), employing two DQN subnetworks for parameter selection and adjustment direction determination while representing states as plan quality scores rather than DVHs. To manage the high dimensionality arising in HNC planning, they manually fixed most of the $141$ potential planning parameters in their case based on dosimetrist expertise, allowing their VTP to learn adjustments for only $8$ unique priority parameters. Similarly, grappling with the high-dimensional action space, \citet{wang2024automating} explored policy-gradient methods with a transformer-based PPO agent for continuous parameter adjustments. Their approach proposed dynamically reducing complexity by adjusting parameters only for the subset of structures currently exhibiting the lowest quality scores. However, this strategy may not accommodate cases where certain OARs have to be strategically compromised to maintain target coverage—a common situation in complex HNC planning scenarios.

Notably, all the aforementioned approaches rely on population-based models trained across diverse patient cohorts. While effective at capturing common anatomical patterns, such models face challenges adapting to patient-specific anatomical changes that necessitate treatment replanning during HNC's fractionated radiotherapy~\citep{caudell2017future, ma2022personalized, maniscalco2023intentional, visak2024assessing}. This limitation, stemming from the under-utilization of intra-patient anatomical consistency, restricts their ability to generate optimally adapted plans tailored to each patient's unique anatomical features and evolving pattern. Motivated by these challenges, we introduce a novel patient-specific RL framework for IMPT replanning in HNC. By developing and optimizing the planning policy to each individual’s unique anatomy and dosimetric constraints from the outset, our approach inherently leverages intra-patient consistency patterns. This personalization is crucial for effectively managing the high-dimensional optimization space and is potentially better suited to addressing the adaptive replanning demands inherent in HNC treatment, overcoming key limitations of existing population-based strategies.

\section{Methods}
\label{sec:methods}

\subsection{Treatment Plan Optimization}

In treatment planning systems (TPS), the patient anatomy (CTVs and OARs) is discretized into voxels. The dose $\Vec{d}_i$ deposited in each voxel $i$ is linearly dependent on adjustable beamlet intensities, represented by the fluence vector $\Vec{x} \in \mathbb{R}^n_+$. This relationship is pre-calculated and stored as a dose influence matrix $\Vec{D}$, where each element $\Vec{D}_{ij}$ quantifies the dose contribution from unit intensity of beamlet $j$ to voxel $i$~\citep{gorissen2022interior}:
\begin{align}
    \label{eq:dij_mat}
    \Vec{d}_i = \sum_j D_{ij} x_j.
\end{align}

The goal of treatment plan optimization is to determine the optimal beamlet intensity vector $\Vec{x}$ that minimizes dose deviations in CTV voxels from their prescribed dose while reducing the dose to OARs, which can be formulated as:
\begin{align}
    \label{eq:plan_optimization_problem}
    \begin{split}
        \min_{\Vec{x} \in \mathbb{R}^n} \quad & \mathcal{L}(\Vec{x}) = \sum_{m=1}^{M} \omega_{\text{CTV}_m} \left\|\Vec{D}_{\text{CTV}_m} \Vec{x} - d_{\text{Rx, CTV}_m} \right\|^2 +  \sum_{k=1}^{K} \omega_{\text{OAR}_k} \left\|\Vec{D}_{\text{OAR}_k} \Vec{x} \right\|^2 \\
        & \text{s.t.} \quad x_j \ge 0, \quad \text{for } j = \{1, \dots, n\}.
    \end{split}
\end{align}
Here, $\mathcal{L}(\Vec{x})$ is the multi-objective treatment planning loss function, where $\Vec{x} \in \mathbb{R}^n$ is the vector of $n$ beamlet intensities to be optimized. Each component $x_j$ represents the intensity or weight of the $j$-th beamlet. The constraint $x_j \ge 0$ explicitly ensures that all beamlet intensities are non-negative. The formulation incorporates $M$ CTVs, where $m \in \{1,\dots,M\}$ indexes each CTV. Similarly, $K$ OARs are considered, where $k \in \{1,\dots, K\}$ indexes each OAR. For the patient cohort in this study, $M \leq 3$ and $K \leq 12$. $\Vec{D}_{\text{CTV}_m}$ and $\Vec{D}_{\text{OAR}_k}$ denote the dose influence matrices for the $m$-th CTV and $k$-th OAR. $d_{\text{Rx, CTV}_m}$ is the prescribed dose vector for the $m$-th CTV.  

The weighting parameters $\omega_{\text{CTV}_m} \geq 0$ and $\omega_{\text{OAR}_k} \geq 0$ serve as treatment planning priorities (TPPs) that balance competing clinical objectives: achieving prescribed dose coverage in the $m$-th CTV while minimizing dose exposure to the $k$-th OAR. While this weighted loss function is a standard proxy for the clinical objective,  it does not directly optimize for the clinical constraints that human planners must satisfy (e.g., specific DVH criteria for various organs listed in Table~\ref{tab:planning_constraints}). In practice, achieving perfect dose homogeneity is not feasible, and different OARs have varying dose tolerances based on their biological characteristics. These clinical realities mean that planners must manually and iteratively adjust the weighting parameters and re-solve the optimization problem until clinical requirements are met--a time-consuming and resource-intensive process.

We automate this critical step by introducing a DRL agent to perform this hyperparameter search. The agent learns to directly optimize for clinical constraint satisfaction, using a reward signal derived from the dose plan's adherence to clinical goals. The DRL agent interacts with a plan optimization engine that serves as the core of our RL environment, minimizing the cost function (Eqn.~\ref{eq:plan_optimization_problem}) with agent-selected priority weights. This integration of dose calculation with RL-based planning optimization streamlines a major bottleneck in the radiotherapy workflow.

\subsection{Reinforcement Learning for Automatic Replanning}

Converting the aforementioned priority tuning during inverse plan optimization into an RL problem offers a solution to the time-consuming, labor-intensive manual adjustment process. We model the priority tuning as a finite Markov Decision Process defined by the tuple ($\mathcal{S}, \mathcal{A}, \mathcal{P}, \mathcal{R}, \gamma$), where:
\begin{itemize}
    \item \textbf{State Space ($\mathcal{S}$)}: Each state $s_t \in \mathcal{S}$ represents the DVHs of the $3$ CTVs and $12$ typical OARs (Table~\ref{tab:planning_constraints}), represented as a normalized $M \times N$ matrix where $M$ corresponds to anatomical structures and $N$ to discretized dose-volume bins.
    \item \textbf{Action Space ($\mathcal{A}$)}: A discrete set of $22$ clinically constrained actions representing priority adjustments for different planning objectives, designed based on clinical experience. Adjustment details are provided in Appendix~\ref{appndx:priority_updates}.
    \item \textbf{Transition Dynamics ($\mathcal{P}$)}: The transition model $\mathcal{P}: \mathcal{S} \times \mathcal{A} \times \mathcal{S} \to [0,1]$ is governed by a treatment optimization engine, with $\mathcal{P}(s_{t+1}|s_t,a_t)$ capturing how priority adjustment actions affect the resulting dose distribution.
    \item \textbf{Reward Function ($\mathcal{R}$)}: The reward function $\mathcal{R}: \mathcal{S} \times \mathcal{A} \to \mathbb{R}$ defines the immediate reward obtained after taking action $a_t$ in state $s_t$. It is defined as $r(s_t, a_t) = \psi(s_t+1) - \psi(s_t)$, where $\psi(\cdot)$ is the plan quality score of the intermediate plan dose, generated by taking action $a_t$. We designed a $150$-point scoring system based on the standardized ProKnow scoring criteria~\citep{nelms2012variation} and institution-specific planning guidelines. It includes a set of dosimetric metrics for the structures of interest. The final score was calculated as the sum of the scores across all these metrics, with higher scores indicating better plan quality. The plot of the scoring function for each dosimetric metric is presented in Appendix~\ref{appndx:pqm}.
    \item \textbf{Discount Factor ($\gamma$)}: $\gamma \in [0,1)$ balances immediate dosimetric improvements with long-term overall plan quality.
\end{itemize}

The goal is to learn an optimal policy $\pi^*$ that dictates the sequence of priority adjustments leading to the highest possible cumulative reward, thereby yielding high-quality treatment plans:
\begin{align*}
    \pi^* = \arg\max_{\pi} \mathbb{E}\left[\sum_{t=0}^{T} \gamma^t r(s_t, a_t) \mid s_0, \pi \right]
\end{align*}
where $\pi: \mathcal{S} \rightarrow \mathcal{A}$ represents a policy that maps DVH states to priority adjustment actions, and $T$ is the planning horizon. The expectation $\mathbb{E}$ is taken over the stochasticity in state transitions under policy $\pi$ starting from initial state $s_0$. 

The optimal policy formulation provides a theoretical objective, but solving for $\pi^*$ directly is challenging in high-dimensional state spaces like those encountered in IMPT replanning. To address this challenge, we implemented two state-of-the-art DRL algorithms that have demonstrated success in complex sequential decision-making tasks.

\paragraph{Deep Q-Network (DQN):} In Q-learning, the objective is to learn the optimal action-value function $Q^*(s, a)$, which represents the maximum expected return achievable by taking action $a$ in state $s$ and following the optimal policy thereafter~\citep{sutton1998reinforcement, watkins1992q}. For proton therapy planning, this corresponds to predicting which priority adjustment ($a$) will yield the greatest long-term improvement in plan quality. The optimal Q-function satisfies the Bellman equation:
\begin{align*}
    Q^*(s, a) = \mathbb{E}_{s' \sim P(\cdot|s,a)} \left[ r(s, a) + \gamma \max{a'} Q^*(s', a') \right]
\end{align*}
DQN approximates this optimal Q-function using a deep neural network $Q(s, a; \Vec{\theta})$ with weights $\Vec{\theta}$. The network is trained by minimizing a loss function that measures the temporal difference (TD) error between the predicted Q-value and the target Q-value~\citep{mnih2015human}. The loss function at iteration $i$ is given by:
\begin{align}
    \label{eq:dqn_loss}
    \mathcal{L}_{\text{DQN}, i}(\Vec{\theta}_i) = \mathbb{E}_{(s, a, r, s') \sim \mathcal{B}} \left[ \left( y_i - Q(s, a; \Vec{\theta}_i) \right)^2 \right]
\end{align}
where $\mathcal{B}$ is a replay buffer storing past experiences $(s, a, r, s')$ (often with prioritized sampling based on TD error), and the target $y_i$ is calculated using a separate target network $Q(s', a'; \Vec{\theta}_i^-)$ with delayed weights $\Vec{\theta}_i^-$, typically updated periodically:
\begin{align*}
    y_i = r + \gamma \max{a'} Q(s', a'; \Vec{\theta}_i^-)
\end{align*}
The DQN agent learns a policy by selecting actions with the highest Q-value for a given state (e.g., using an $\epsilon$-greedy strategy to balance exploration and exploitation).

\paragraph{Proximal Policy Optimization (PPO):} While DQN learns Q-values of state-action pairs and derives policies through action selection mechanisms, PPO takes a more direct approach. PPO is an actor-critic policy gradient algorithm that explicitly learns a policy $\pi((a|s); \Vec{\theta}_{\Vec{p}})$ parameterized by weights $\Vec{\theta}_{\Vec{p}}$ and a value function $V(s;\Vec{\theta}_{\Vec{\nu}})$ parameterized by weights $\Vec{\theta}_{\Vec{\nu}}$. The policy (actor) determines actions based on the current state, while the value function (critic) approximates the value of the current state, providing a baseline for evaluating the actor's performance. Policy gradient methods update the actor's parameters by following the gradient of an objective function that aims to maximize the expected return~\citep{schulman2017proximal}. To ensure stable training, PPO employs a clipped surrogate objective function that prevents excessively large policy updates by maintaining the new policy within a trust region of the old policy. The objective function for the actor is defined as:
\begin{align*}
    L_{\text{actor}} = \mathbb{E}_{(s, a) \sim \pi_{\Vec{\theta}_{\Vec{p}_{\text{old}}}}} \Bigg[ \min\bigg(\frac{\pi_{\Vec{\theta}_{\Vec{p}}}(a|s)}{\pi_{\Vec{\theta}_{\Vec{p}_{\text{old}}}}(a|s)} \hat{A}_t(s,a), \text{clip}\left( \frac{\pi_{\Vec{\theta}_{\Vec{p}}}(a|s)}{\pi_{\Vec{\theta}_{\Vec{p}_{\text{old}}}}(a|s)}, 1-\epsilon, 1+\epsilon \right) \hat{A}_t(s,a) \bigg) \Bigg]
\end{align*}
where $\pi_{\Vec{\theta}_{\Vec{p}_{\text{old}}}}$ is the policy from the previous iteration, $\pi_{\Vec{\theta}_{\Vec{p}}}$ is the current policy, $\epsilon$ is a hyperparameter controlling the clipping range, and $\hat{A}_t(s,a)$ is the generalized advantage estimate (GAE)~\citep{schulman2015high}:
\begin{align}
    \label{eq:gae}
    \hat{A}_t(s_t,a_t) &= \sum_{l=0}^{T-t-1} (\gamma\lambda)^l \bigg( r_{t+l} + \gamma V_{\Vec{\theta}_{\Vec{\nu}}}(s_{t+l+1}) - V_{\Vec{\theta}_{\Vec{\nu}}}(s_{t+l}) \bigg)
\end{align}
where $\gamma \in [0,1)$ is the discount factor, $\lambda \in [0,1]$ controls the bias-variance tradeoff, and $V_{\Vec{\theta}_{\Vec{\nu}}}$ represents the value function approximation. GAE estimates how much better an action is compared to the average action in a given state.

Simultaneously, the critic is trained to accurately estimate the state value by minimizing the mean squared error between its predictions and the actual discounted return:
\begin{align*}
    L_{\text{critic}} = \mathbb{E}_{(s_t, R_t) \sim \mathcal{B}} \left[ \left( V_{\Vec{\theta}_{\Vec{\nu}}}(s_t) - R_t \right)^2 \right]
\end{align*}
where $R_t = \sum_{k=0}^{T-t}\gamma^k r_{t+k}$ is the discounted return and $\mathcal{B}$ is the experience replay of the current policy.

In essence, PPO aims to find optimal policy and value function parameters by iteratively minimizing a combined objective function that includes both the actor loss and the critic loss, thereby approximately solving the minimization problem given by:
\begin{align}
    \label{eq:ppo_loss}
    \mathcal{L}_{\text{PPO}} = \min_{\Vec{\theta}_{\Vec{p}}, \Vec{\theta}_{\Vec{\nu}}} \mathbb{E}_{(s, a) \sim \mathcal{B}} \left[ -L_{\text{actor}}(\Vec{\theta}_{\Vec{p}}) + c_1 L_{\text{critic}}(\Vec{\theta}_{\Vec{\nu}}) - c_2 \mathbb{E}_s[H(\pi_{\Vec{\theta}_{\Vec{p}}}(\cdot|s))] \right]
\end{align}
where $c_1$ and $c_2$ are coefficients controlling the relative importance of the value function loss and the entropy bonus $H(\cdot)$, respectively. The entropy bonus encourages exploration.

\section{Patient Cohort}
\label{sec:cohort_overview}

\subsection{Cohort Selection} 

We validated our approach using a retrospective cohort of $18$ HNC patients treated with IMPT at the Emory Proton Therapy Center. Eight patients requiring replanning due to substantial anatomical changes were used for patient-specific RL training, while the remaining 10 patients served as training data for population-based RL baselines. For each patient, an initial planning CT (pCT) was acquired $1-2$ weeks before the treatment course, and a replanning CT (rpCT) was obtained when significant anatomical changes were observed and warranted replanning. Performance evaluation used all eight rpCT cases for both RL approaches. These cases were selected for their varied tumor characteristics and spatial relationships to surrounding OARs, all involving three CTVs representing challenging planning scenarios. While the primary CTV (CTV1) for all cases was prescribed $70$~\si{\rbe} over $35$ fractions, the prescription doses for the secondary (CTV2) and tertiary (CTV3) targets varied per case (Table~\ref{tab:treatment_volumes}). This prescription heterogeneity, combined with substantial inter-patient anatomical variation, poses a major challenge for automated treatment planning in HNC, particularly for population-based models struggling to generalize across heterogeneous clinical objectives. Patient demographics and staging are detailed in Table~\ref{tab:patient_demographics} (Appendix~\ref{appndx:cohort}).

\begin{table}[t]
    \centering
    \caption{\textbf{Treatment Parameters and CTV Volumetric Changes}. Prescription doses, replanning frequency, and CTV volumes (\si{cc}) on planning CT (pCT) and replan CTs (rpCT) for primary CTV (CTV1), secondary CTV (CTV2), and tertiary CTV (CTV3). All patients received 35 fractions. NA indicates no second replan was performed.}
    \label{tab:treatment_volumes}
    \resizebox{\textwidth}{!}{
        \begin{tabular}{c SSS c SSSSSS SSSSSS}
            \toprule
            \multirow{3}{*}{\textbf{Case ID}} & 
            \multicolumn{3}{c}{\textbf{Prescription (\si{\rbe})}} & 
            \multirow{3}{*}{\textbf{\shortstack{No. of \\ Replans}}} &
            \multicolumn{3}{c}{\textbf{Volume on pCT (\si{cc})}} & 
            \multicolumn{3}{c}{\textbf{Volume on \nth{1} rpCT (\si{cc})}} & 
            \multicolumn{3}{c}{\textbf{Volume on \nth{2} rpCT (\si{cc})}} \\
            \cmidrule(lr){2-4} \cmidrule(lr){6-8} \cmidrule(lr){9-11} \cmidrule(lr){12-14}
            & {CTV1} & {CTV2} & {CTV3} & & {CTV1} & {CTV2} & {CTV3} & {CTV1} & {CTV2} & {CTV3} & {CTV1} & {CTV2} & {CTV3} \\
            \midrule
            P1 & 70.00 & 59.85 & 53.90 & 2 & 182.58 & 441.98 & 199.84 & 150.30 & 386.54 & 196.90 & 133.45 & 362.06 & 182.75 \\
            P2 & 70.00 & 63.00 & 56.00 & 1 & 182.65 & 419.51 &  55.78 & 172.37 & 403.60 &  64.11 & {NA} & {NA} & {NA} \\
            P3 & 70.00 & 59.85 & 53.90 & 1 & 206.38 & 522.20 &  40.53 & 184.04 & 453.84 &  35.69 & {NA} & {NA} & {NA} \\
            P4 & 70.00 & 63.00 & 56.00 & 2 &  84.40 & 122.69 & 219.46 &  69.63 & 106.29 & 217.84 &  65.78 & 102.78 & 195.47 \\
            P5 & 70.00 & 63.00 & 56.00 & 1 & 211.17 & 362.35 & 485.46 & 194.52 & 346.14 & 476.32 & {NA} & {NA} & {NA} \\
            P6 & 70.00 & 60.20 & 53.90 & 2 & 242.93 & 246.37 &  50.93 & 297.74 & 253.30 &  52.11 & 180.45 & 232.65 &  55.72 \\
            P7 & 70.00 & 63.00 & 56.00 & 2 & 122.19 & 263.02 & 440.24 & 132.33 & 267.06 & 442.82 & 120.17 & 255.05 & 434.99 \\
            P8 & 70.00 & 59.85 & 53.90 & 2 &  89.09 & 229.30 & 185.22 &  83.10 & 213.64 & 171.83 &  77.36 & 213.06 & 174.36 \\
            \bottomrule
        \end{tabular}
    }
\end{table}

\subsection{Data Extraction} 

Relevant patient data, including the pCT and rpCT, and their corresponding CTV and OAR contours, were retrieved from an internal database in DICOM format. All the contours were manually delineated by attending physicians for treatment planning or replanning of the actual treatments. These DICOM files of CT images and contours were imported into the open-source treatment planning toolkit matRad~\citep{wieser2017development} to precalculate dose influence matrices ($\Vec{D}$) required for plan optimization, using the same beam arrangements (e.g., isocenter location, number of beams, and beam angles) as those used in the actual treatments. The dose matrix calculated from the pCT and its associated contours was used for training the DRL agents, while the matrix calculated from the rpCT and corresponding contours was used for testing. 

To train a patient-specific agent capable of adapting treatment plans to anatomical changes, we generated augmented anatomies by simulating tumor variations. In this proof-of-concept study, we simulated only two variation scenarios, i.e., tumor progression and regression, by expanding the original CTV contours on the pCT by $2$~\si{mm} or shrinking them by $3$~\si{mm}, respectively. These variations not only changed the CTV volumes but also altered their spatial relationships with surrounding OARs, increasing the diversity of the patient-specific training dataset in terms of anatomy and planning complexity. The pCT images, along with each set of modified contours, were also imported into matRad to calculate the corresponding dose influence matrices for the simulated anatomy variations.

\section{Experiments \& Results} 
\label{sec:exp_results}

In this section, we present the results of our experiments and evaluate the effectiveness of RL-based planners for automated IMPT replanning in HNC treatment. 

\subsection{Experimental Setup}

For each patient, both DQN and PPO agents were trained using the patient’s original anatomy from the pCT and two augmented anatomical variations. The state representation included the normalized DVH curves for all structures \big(dim=$(15, 100)$\big), while the discrete action space consisted of $22$ priority adjustments for CTVs and OARs. Each agent was trained for $100$ episodes, with each episode consisting of up to $15$ priority adjustment steps or terminating early if the maximum plan quality score of $150$ was achieved. Algorithm~\ref{alg:drl_impt_replan} in Appendix~\ref{appndx:alg_protonRL} provides a detailed overview of this DRL-based replanning process, illustrating the integration of priority tuning and plan optimization within our experimental framework. Details of network specifications and hyperparameters are provided in Appendix~\ref{appndx:network_details}. 

The trained agents were then tested on the patient’s new anatomy captured in the rpCT. Specifically, starting from an initial default priority set $\Vec{\omega}_0$, the agents performed priority tuning following the same episodic framework described above, where each step involved selecting a tuning action based on the intermediate plan's quality, followed by inverse plan optimization. The number of tuning attempts (horizon length) in this process can be extended as needed, bounded only by computational cost. As a comparative baseline, a human planner also generated a new manual treatment plan for each case, employing manual priority tuning. The performance of the agents was assessed by comparing the plan quality scores and dosimetric metrics of the agent-generated plans with those of the manually created plans. 

To facilitate the interaction between the RL agents and the treatment planning process, we developed a custom OpenAI Gymnasium-compatible environment that simulates the treatment planning workflow. At each time step, the environment receives the agent-selected action at $a_t$ and applies it to modify the current priority weights $\Vec{\omega}$. Using these updated priorities along with the dose influence matrices $\Vec{D}_{\text{CTV}m}$ and $\Vec{D}_{\text{OAR}_k}$ for each CTV $m$ and OAR $k$, the environment solves the optimization problem (Eqn.~\ref{eq:plan_optimization_problem}) via projected gradient descent~\citep{fu2023distributed, ghobadi2012automated, nocedal1999numerical} to update the beamlet weights $\Vec{x}$ and generate a new dose distribution $\Vec{d}$ (Eqn.~\ref{eq:dij_mat}). The resulting dose distribution is evaluated against clinical constraints (Table~\ref{tab:planning_constraints}) to compute the plan score and reward (Appendix~\ref{appndx:pqm}). Each action $a \in \{0,1,\dots,21\}$ updates priority weights $\Vec{\omega}$ according to
\begin{align}
    \label{eq:priority_update_rule}
    \Vec{\omega}_{(t+1)} = \big[ \Vec{\omega}_{(t)} + \Delta_a \big]_{0}^{1}, \quad t\in[0,14],
\end{align}
where $[x]_0^1 \triangleq \min(\max(x, 0), 1)$ ensures weights remain within valid bounds. The details of the priority adjustment $\Delta_a$ are included in the Appendix~\ref{appndx:priority_updates}. All experiments were conducted on a workstation equipped with an NVIDIA RTX $5000$ Ada Generation GPU with $32$~\si{GB} of memory.

\begin{table}[t]
    \centering
    \caption{\textbf{Plan quality scores on \nth{1} replanning CT ($0\text{--}150$ scale; higher preferred).} Comparison of treatment plans generated using the initial default priority set (referred to as \textit{initial}), manually created plans (\textit{manual}), and plans automatically generated by both population-based(\textit{-popn}) and patient-specific DQN and PPO agents. Bold values denote the highest score for each patient.}
    \label{tab:scores}
    \resizebox{\textwidth}{!}{%
    \begin{tabular}{cccccccc}
        \toprule
        \textbf{Case ID} & \textbf{Initial} & \textbf{Manual} & \textbf{DQN\textit{-popn}} & \textbf{DQN} & \textbf{PPO\textit{-popn}}  & \textbf{PPO} \\
        \midrule
        P1 & 122.56 & 130.85 & 130.40 & 131.05 & 137.01  & \textbf{137.88} \\
        P2 & 83.14 & 132.09 &  135.33 & 136.20 & 135.64  & \textbf{138.05} \\
        P3 & 132.35 & 140.44 & 144.17  & 147.25 &  147.31 & \textbf{147.52} \\
        P4 & 132.14 & 143.94 & 148.38  & 146.10 & 146.38  & \textbf{148.77} \\
        P5 & 132.96 & 138.67 &  138.41 & 138.32 & 138.30  & \textbf{141.50} \\
        P6 & 125.65 & 133.03 & 128.87  & \textbf{135.24} & 135.18  & 135.18 \\
        P7 & 108.67 & 132.53 & 126.27  & 141.03 & 136.51  & \textbf{141.35} \\
        P8 & 128.75 & 139.03 & 124.31  & 141.53 & 137.06  & \textbf{141.78} \\
        \midrule
        \textbf{mean $\pm$ std} & $120.78 \pm 17.18$ & $136.32 \pm 4.79$ & $134.52 \pm 8.64$ & $139.59 \pm 5.50$ & $139.17 \pm 4.83$ & $\mathbf{141.50 \pm 4.69}$ \\
        \bottomrule
    \end{tabular}
    }
\end{table}

\subsection{Plan Quality Comparisons}

The same $150$-point scoring system used to calculate rewards during DRL training was employed to quantify the plan quality of treatment plans generated for each patient's new anatomy captured in the rpCT. To provide comprehensive benchmarking, we evaluated our patient-specific approach against both manual planning and population-based RL baselines. The population-based agents were trained on the initial planning CTs of the remaining $10$ patients in our cohort using identical architecture, hyperparameters, and training setup, consistent with conventional clinical practice where models are trained on population data.

The resulting plan quality scores on the \nth{1} replanning CT are presented in Table~\ref{tab:scores}, with detailed dosimetric metrics summarized in Appendix~\ref{appndx:complete_results}. All RL approaches effectively adjusted planning objective priorities, significantly improving plan quality compared to initial plans generated using default priority sets. Crucially, patient-specific PPO achieved the highest mean quality score ($141.50 \pm 4.69$), outperforming manual plans ($136.32 \pm 4.79$, $p < 0.01$ via paired \textit{t}-test), population-based PPO ($139.17 \pm 4.83$), patient-specific DQN($139.59 \pm 5.50$), and population-based DQN ($134.52 \pm 8.64$). Furthermore, the patient-specific PPO-generated plans exhibited lower score variability ($\sigma = 4.69$), indicating enhanced consistency and reliability in adapting to anatomical changes compared to all other methods.

Overall, patient-specific agents matched or exceeded manual plans in all cases, with PPO achieving the highest score in $7$ out of $8$ patients. Beyond manual benchmarks, patient-specific agents significantly outperformed population-based agents in $14$ of the $16$ comparisons (across all patients and algorithms), underscoring the clear advantages of personalized adaptation over generalized models, especially when handling unique anatomical changes.

\begin{figure}
    \centering
    \includegraphics[width=0.8\linewidth]{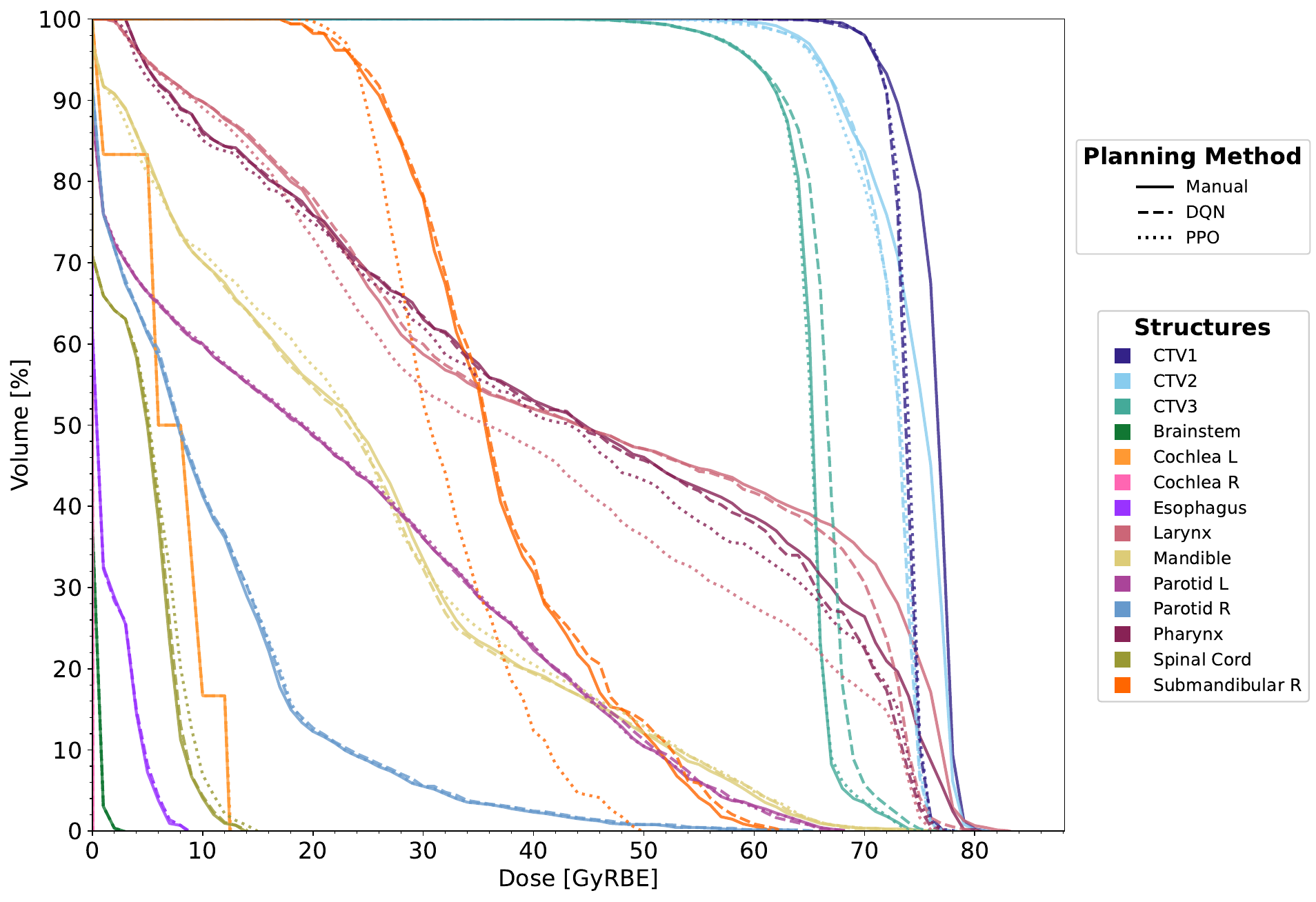}
    \caption{\textbf{DVH comparison for patient P4's \nth{1} replanning CT:} Manual replans (solid lines), patient-specific DQN (dashed lines), and patient-specific PPO (dotted lines). Complete DVH curves and dosimetric analysis for all patients' \nth{1} rpCTs are provided in Appendix~\ref{appndx:complete_results}.}
   \label{fig:p4_dvh_curves}
\end{figure}

\subsection{Dosimetric Analysis}

 Detailed dosimetric metrics for each structure on the \nth{1} replanning CT, comparing manually generated plans with those from patient-specific DQN and PPO agents, are presented in Table~\ref{tab:full_results} (P1-P5) and Table~\ref{tab:additional_dosimetric_results} (P6-P8, Appendix~\ref{appndx:complete_results}). Clinically acceptable target dose coverage was achieved in all the three plans for every case, with the coverage of primary CTV ($V_{d_{Rx,\text{CTV}_1}}$) no less than $97.97\%$. In terms of dose homogeneity inside the primary CTV (CTV1), both RL agents reduced hot spots in three patients while maintaining excellent coverage. Most notably, in patient P4, DQN and PPO reduced the maximum dose within CTV1 by $3.41$~\si{\rbe} and $2.65$~\si{\rbe}, respectively, compared to the manual plan. PPO demonstrated superior OAR sparing in most patients, reducing mean dose to the larynx (LAR) \big($36.15 \pm 3.38$~\si{\rbe} (PPO) vs. $38.48 \pm 5.76$~\si{\rbe} (Manual)\big) and to the bilateral parotids (PARL and PARR) \big($19.81 \pm 5.86$~\si{\rbe} (PPO) vs. $22.00 \pm 8.66$~\si{\rbe} (Manual)\big), compared to manual replanning. The effectiveness of PPO was particularly evident in case P5, where it slightly compromised the dose coverage of the tertiary CTV (CTV3, $97.53\%$) to substantially reduce the dose to the right submandibular gland (SMGR), bringing its mean dose closer $37.60$~\si{\rbe} to the dose tolerance ($35$~\si{\rbe}) and resulting in a significantly higher plan score. In contrast, the manual and DQN-generated plans for this case substantially exceeded the dose tolerance of the SMGR, with mean doses of $67.28$~\si{\rbe} and $70.89$~\si{\rbe}, respectively. Figure~\ref{fig:p4_dvh_curves} compares the DVH curves for patient P4, illustrating results from manually generated plans and those produced by patient-specific RL agents (DQN and PPO).

 \begin{table}[t]
     \centering
     \caption{\textbf{Summary of dosimetric performance across patients P1-P5 on \nth{1} rpCT:} Manual (M), patient-specific DQN (Q), and patient-specific PPO (P). All dose metrics ($D_{\text{0\%/0.03\si{cc}/mean}}$) in~\si{\rbe}. Bold values indicate superior dosimetry outcomes. Results for P6-P8 in Table~\ref{tab:additional_dosimetric_results} (Appendix~\ref{appndx:complete_results}).}
     \label{tab:full_results}
     \resizebox{\textwidth}{!}{%
        \begin{tabular}{ll*{15}{c}}
            \toprule
            \multirow{2}{*}{\textbf{Structure}} & \multirow{2}{*}{\textbf{Metric}} & 
            \multicolumn{3}{c}{\textbf{P1}} & \multicolumn{3}{c}{\textbf{P2}} & 
            \multicolumn{3}{c}{\textbf{P3}} & \multicolumn{3}{c}{\textbf{P4}} & 
            \multicolumn{3}{c}{\textbf{P5}} \\
            \cmidrule(lr){3-5} \cmidrule(lr){6-8} \cmidrule(lr){9-11} \cmidrule(lr){12-14} \cmidrule(lr){15-17}
            & & {M} & {Q} & {P} & {M} & {Q} & {P} & {M} & {Q} & {P} & {M} & {Q} & {P} & {M} & {Q} & {P} \\
            \midrule
            \multirow{2}{*}{CTV1} 
            & $V_{d_{Rx,\text{CTV}_1}} \geq 98\%$ & 97.99 & \textbf{98.01} & 97.99 & 97.99 & \textbf{98.01} & 97.99 & 97.99 & \textbf{98.00} & 97.99 & \textbf{98.01} & 97.97 & \textbf{98.01} & 97.99 & 97.99 & \textbf{98.00}\\
            & $D_{0\%} \leq 77$ & 81.86 & \textbf{81.02} & 82.24 & \textbf{82.30} & 83.03 & 83.61 & 81.52 & 81.00 & \textbf{79.68} & 81.23 & \textbf{77.82} & 78.58 & \textbf{77.26} & 80.82 & 77.82 \\
            \midrule
            CTV2 & $V_{d_{Rx,\text{CTV}_2}} \geq 98\%$ & 97.97 & 97.99 & \textbf{98.56} & 97.94 & \textbf{98.27} & 98.04 & \textbf{98.39} & \textbf{98.39} & 98.29 & \textbf{98.55} & 98.14 & 98.19 & \textbf{99.51} & 98.50 & 99.19 \\
            CTV3 & $V_{d_{Rx,\text{CTV}_3}} \geq 98\%$ & 97.84 & 98.03 & \textbf{99.00} & 97.76 & \textbf{98.27} & 98.18 & 98.18 & \textbf{98.33} & 98.02 & 98.02 & 97.94 & \textbf{98.03} & \textbf{99.12} & 97.98 & 97.53 \\
            \midrule
            BRS & $D_{\text{0.03cc}} \leq 30$ & \textbf{23.29} & 24.01 & 24.17 & 19.59 & \textbf{19.58} & 20.04 & 10.69 & \textbf{10.55} & 10.77 & 2.87 & \textbf{2.82} & 2.85 & 13.69 & \textbf{12.48} & 14.07 \\
            SC  & $D_{\text{0.03cc}} \leq 30$ & \textbf{37.63} & 37.67 & 40.60 & \textbf{22.63} & 22.70 & 22.85 & \textbf{31.81} & 32.35 & 32.17 & \textbf{13.85} & 14.04 & 14.98 & 30.58 & \textbf{30.23} & 32.06 \\
            MAN & $V_{70\si{\rbe}} \leq 10\%$ & 1.73 & \textbf{0.58} & 2.18 & 1.29 & \textbf{1.07} & 1.51 & 0.13 &  \textbf{0.00} & 1.28 & 0.22 & \textbf{0.09} & \textbf{0.09} & \textbf{0.08} & 0.67 & 0.20 \\
            LAR & $D_{\text{mean}} \leq 40$ & 43.76 & 43.84 & \textbf{40.15} & 36.51 & \textbf{36.48} & 36.76 & \textbf{34.76} & 34.84 & 34.91 & 45.29 & 44.51 & \textbf{39.29} & 32.10 & 32.74 & \textbf{30.86} \\
            PHY & $D_{\text{mean}} \leq 50$ & \textbf{0.00} & \textbf{0.00} & \textbf{0.00} & 49.13 & \textbf{48.95} & 49.41 & \textbf{0.00} & \textbf{0.00} & \textbf{0.00} & 44.00 & 43.17 & \textbf{42.13} & 37.73 & 38.98 & \textbf{35.12} \\
            PARL & $D_{\text{mean}} \leq 26$ & \textbf{14.43} & 14.57 & 15.56 & 26.00 & 26.00 & \textbf{22.74} & 20.39 & \textbf{20.19} & 20.52 & \textbf{21.76} & 21.79 & 21.86 & 10.04 & \textbf{9.63} & 10.12 \\
            PARR & $D_{\text{mean}} \leq 26$ & 34.60 & 34.50 & \textbf{23.86} & 28.41 & \textbf{25.15} & 25.81 & 32.68 & 25.97 & \textbf{25.71} & \textbf{10.08} & 10.28 & 10.19 & \textbf{21.56} & 21.81 & 21.69 \\
            COCHL & $D_{\text{mean}} \leq 35$ & 4.03 & 4.15 & \textbf{3.96} & 39.34 & 39.82 & \textbf{33.55} & \textbf{0.02} & \textbf{0.02} & \textbf{0.02} & \textbf{6.92} & 6.93 & 6.97 & \textbf{0.00} & \textbf{0.00} & \textbf{0.00} \\
            COCHR & $D_{\text{mean}} \leq 35$ & 2.09 & \textbf{2.07} & 2.42 & \textbf{8.44} & 8.47 & 8.56 & \textbf{0.84} & 0.88 & \textbf{0.84} & \textbf{0.02} & \textbf{0.02} & \textbf{0.02} & \textbf{0.00} & \textbf{0.00} & \textbf{0.00} \\
            SMGL & $D_{\text{mean}} \leq 35$ & \textbf{59.73} & 60.59 & 60.07 & 72.57 & \textbf{72.11} & 72.68 & \textbf{0.00} & \textbf{0.00} & \textbf{0.00} & \textbf{0.00} & \textbf{0.00} & \textbf{0.00} & \textbf{68.14} & 71.57 & 68.15 \\
            SMGR & $D_{\text{mean}} \leq 35$ & \textbf{0.00} & \textbf{0.00} & \textbf{0.00} & 72.83 & \textbf{72.51} & 73.06 & \textbf{0.00} & \textbf{0.00} & \textbf{0.00} & 36.87 & 37.30 & \textbf{31.98} & 67.28 & 70.89 & \textbf{37.60} \\
            ESO & $D_{\text{mean}} \leq 40$ & \textbf{11.09} & \textbf{11.09} & 11.57 & \textbf{6.23} & 6.26 & 6.24 & \textbf{13.26} & 13.35 & 13.31 & \textbf{1.36} & 1.38 & 1.38 & 6.41 & \textbf{5.85} & 6.69 \\
            \midrule
            \multicolumn{2}{l}{\textbf{Plan Score (max=150):}} & 130.85 & 131.05 & \textbf{137.88}& 132.09 & 136.20 & \textbf{138.05} & 140.44 & 147.25 & \textbf{147.52} & 143.94 & 146.10 & \textbf{148.77} & 138.67 & 138.32 & \textbf{141.50} \\
            \bottomrule
        \end{tabular}%
     }
    \begin{minipage}{0.9\textwidth}
        \footnotesize
        \vspace{0.5em}
        \textit{Note}: OAR abbreviations - BRS: Brainstem, SC: Spinal Cord, MAN: Mandible, LAR: Larynx, PHY: Pharynx, PARL/PARR: Left/Right Parotid, COCHL/COCHR: Left/Right Cochlea, SMLG/SMGR: Left/Right Submandibular Gland, ESO: Esophagus.
    \end{minipage}
 \end{table}

\subsection{Adaptation to Inter-fractional Changes}

To further assess the robustness of our approaches in handling rapidly evolving anatomies, we extended our comparison to include the second replanning CT for cases in our evaluation cohort that underwent multiple replanning sessions during the treatment course (P1, P4, P6, P7, P8). As shown in Table~\ref{tab:replan_scores}, patient-specific RL demonstrated superior adaptation capabilities, achieving better performance than population-based baselines in $9$ out of $10$ metric comparisons (DQN: $4/5$ cases; PPO: $5/5$ cases). Patient-specific DQN, in particular, achieved a higher mean score of $138.43 \pm 3.91$ compared to $127.78 \pm 12.14$ for the population-based DQN baseline, while showing markedly lower variability. Although population-based approaches showed efficacy in handling certain cases with common anatomical patterns (e.g., P4 with DQN), the patient-specific approach consistently demonstrated better adaptation to the unique inter-fractional changes of individual patients.

\begin{table}[ht]
    \centering
    \caption{\textbf{Population-based(\textit{-popn}) vs. patient-specific DRL performance on \nth{2} replanning CT:} Comparison of DQN and PPO approaches for patients requiring multiple replanning sessions. Plan quality scores ($0\text{--}150$ scale; higher preferred).}
    \label{tab:replan_scores}
    \resizebox{0.8\textwidth}{!}{%
        \begin{tabular}{cccccc}
            \toprule
            \textbf{Case ID}  & \textbf{DQN\textit{-popn}} & \textbf{DQN} & \textbf{PPO\textit{-popn}} & \textbf{PPO} \\
            \midrule
            P1 & 124.95 & 133.33 & 126.78 & \textbf{133.64} \\
            P4 & \textbf{146.23} & 143.13 & 145.11 & 145.88 \\
            P6 & 132.55 & 135.69 & 135.16 & \textbf{136.54} \\
            P7 & 115.27 & 140.57 & 135.21 & \textbf{141.11} \\
            P8 & 119.92 & \textbf{139.42} & 130.26 & 138.92 \\
            \midrule
            \textbf{mean $\pm$ std} & $127.78 \pm 12.14$ & $138.43 \pm 3.91$ & $134.50 \pm 6.91$ & $\mathbf{139.22 \pm 4.65}$ \\
            \bottomrule
        \end{tabular}
    }
\end{table}

\section{Discussion} 
\label{sec:discuss}

Our study demonstrates that patient-specific RL-based automated IMPT replanning, particularly using PPO, consistently generates treatment plans of higher quality for HNC patients experiencing anatomical changes during the treatment course, compared to both manually generated plans and population-based RL approaches. These improvements have several important implications for adaptive proton therapy workflows.

\subsection{Clinical Significance}

The marked improvement in plan quality scores across all patients indicates that patient-specific RL-based automated replanning can standardize high-quality treatment planning tailored to individual anatomical variations. This personalized approach may lead to reduced treatment-related toxicities while maintaining effective tumor control. For example, the reduction in parotid gland dose has been correlated with decreased risk of xerostomia, with studies suggesting that each $1$~\si{\rbe} reduction in mean dose translates to approximately $4\%$ reduction in xerostomia risk~\citep{castelli2015impact, chao2001prospective}. As shown in Table~\ref{tab:full_results}, patients P1 and P2 benefited from substantial parotid sparing (reductions of $10.7$~\si{\rbe} and $3.3$~\si{\rbe}, respectively), while patients P1 and P4 experienced notable laryngeal sparing (reductions of $3.6$~\si{\rbe} and $6.0$~\si{\rbe}).

Importantly, these dosimetric improvements were achieved without compromising target coverage, underscoring the ability of patient-specific RL methods to address the geometric and dosimetric complexities unique to each patient's HNC anatomy and balance complex trade-offs to achieve clinically acceptable treatment plans. The superior performance of patient-specific agents over population-based approaches ($14/16$ comparisons in Table~\ref{tab:scores}) demonstrates that individualized adaptation is crucial for handling the substantial inter-patient anatomical variability characteristic of HNC. The proposed patient-specific RL-based automated replanning approach offers a promising solution for efficient offline adaptation and paves the way for personalized online adaptive proton therapy.

\subsection{RL-based Planning and Patient-specific Benefits}

Patient-specific RL algorithms outperformed both manual replanning and population-based counterparts, with PPO consistently delivering superior plans compared to DQN. This finding aligns with recent literature suggesting policy-based methods like PPO effectively navigate the complex reward landscapes common in radiotherapy planning~\citep{wang2024automating}. PPO's strategy of performing small, incremental parameter updates helps prevent detrimental large policy shifts, a feature particularly well-suited to patient-specific radiotherapy optimization where individual anatomical nuances require careful navigation.

The advantage of patient-specific adaptation becomes evident when analyzing replanning outcomes relative to individual anatomical features (Table~\ref{tab:treatment_volumes}). On the \nth{1} replanning CT (Table~\ref{tab:scores}), patient-specific PPO achieved the highest plan scores in $7/8$ cases, outperforming manual replans across diverse target volumes (rpCT CTV1 range: $69.63 \text{--} 297.74$~\si{cc}) and volumetric changes between pCT and rpCT (e.g., P2 CTV3: $+15.0\%$ expansion). This superiority is highlighted by its achievement of the highest absolute score ($148.77$) for the patient with the smallest CTV1/CTV2 (P4) and clinically meaningful gains for large-volume cases over manual plans (e.g., patient P3: $+7.08$; P5: $+2.83$). Furthermore, the lower score variability of patient-specific approaches ($\sigma = 4.69$ for PPO) suggests more consistent adaptation to each patient's unique anatomical constraints.

This performance gap widens on the \nth{2} replanning CT (Table~\ref{tab:replan_scores}), where the benefits of patient-specific DRL become even more pronounced. Both patient-specific DQN and PPO agents consistently and significantly outperformed their population-based counterparts. Patient-specific PPO, in particular, achieved the highest overall mean score (PPO: $139.22 \pm 4.65$ vs. PPO-popn: $134.50 \pm 6.91$). In contrast, the high score variability observed in population-based approaches ($\sigma = 12.14$ for DQN-popn) demonstrates that these models struggle to generalize and consistently adapt to the diverse and dynamic anatomical changes. This confirms that the benefits of patient-specific training persist and amplify as anatomical changes accumulate throughout the treatment course.

These findings collectively demonstrate that patient-specific DRL delivers significant dosimetric advantages over population-averaged approaches by learning and adapting to individual patient characteristics. By consistently delivering high-quality plans tailored to each patient's evolving anatomy, this framework holds considerable promise for improving adaptive radiotherapy outcomes in HNC treatment.

\subsection{Limitations and Future Directions}

While our findings demonstrate the feasibility of deep reinforcement learning (DRL) for adaptive IMPT replanning, important limitations should be acknowledged alongside opportunities for future work. First, the retrospective evaluation was conducted on a limited cohort of eight patients. However, this cohort is representative of challenging HNC cases with three CTVs of different prescription levels and common anatomical change patterns. Future work should apply our patient-specific approach to a larger, more diverse cohort of patients with varying CTV configurations, rigorously assessing the framework's adaptability across different anatomical presentations. The demonstrated superiority of our patient-specific DRL agents over manual plans further motivates comprehensive evaluation in prospective studies.

Second, the current framework’s reliance on simulated anatomical changes, while methodologically necessary for controlled RL development, necessitates validation with real longitudinal anatomical changes observed during actual treatment courses. Simulated target volume expansions and contractions may not fully capture the complex, heterogeneous tissue deformations, weight loss patterns, tumor shrinkage dynamics, and normal tissue responses that occur during radiotherapy. Future work should incorporate actual or additional anatomical change scenarios for training to enhance the performance of the DRL agents further. 

Finally, methodological advancements to the RL agent itself could yield further dosimetric improvements. While our current approach focuses on clinically interpretable discrete priority adjustments, exploring continuous action spaces could potentially allow for finer control over planning parameters and enable more nuanced treatment planning. Exploring more sophisticated state representations may further reduce computational demands and enhance plan quality. To address the computational overhead associated with patient-specific training, transfer learning techniques might be needed. Such techniques could facilitate the development of readily deployable models, requiring only minimal fine-tuning for new patients. 

\section{Conclusion}
Reinforcement learning, particularly PPO-based approaches, offers a compelling approach to automated replanning in HNC IMPT. The patient-specific nature of our RL framework enables tailored optimization strategies that adapt to the unique anatomical and dosimetric challenges of each patient. Our findings demonstrate a consistent generation of superior treatment plans compared to manual planning, potentially reducing planning time and improving plan quality. These results suggest that RL-based solutions can significantly enhance IMPT workflows, ultimately benefiting cancer patients through reduced toxicities and effective tumor control.

\acks{This project is supported by the National Institute of Health under Award Numbers\\ R01DE033512 and R37CA272755. We thank Dr. David S. Yu, Dr. Mark McDonald, and Dr. Ralph Weichselbaum for providing valuable clinical perspectives that informed this study.}

\bibliography{references}

\begin{thebibliography}{42}
\providecommand{\natexlab}[1]{#1}
\providecommand{\url}[1]{\texttt{#1}}
\expandafter\ifx\csname urlstyle\endcsname\relax
  \providecommand{\doi}[1]{doi: #1}\else
  \providecommand{\doi}{doi: \begingroup \urlstyle{rm}\Url}\fi

\bibitem[Barker~Jr et~al.(2004)Barker~Jr, Garden, Ang, O'Daniel, Wang, Court, Morrison, Rosenthal, Chao, Tucker, et~al.]{barker2004quantification}
Jerry~L Barker~Jr, Adam~S Garden, K~Kian Ang, Jennifer~C O'Daniel, He~Wang, Laurence~E Court, William~H Morrison, David~I Rosenthal, KS~Clifford Chao, Susan~L Tucker, et~al.
\newblock Quantification of volumetric and geometric changes occurring during fractionated radiotherapy for head-and-neck cancer using an integrated ct/linear accelerator system.
\newblock \emph{International Journal of Radiation Oncology, Biology, Physics}, 59\penalty0 (4):\penalty0 960--970, 2004.

\bibitem[Bobi{\'c} et~al.(2023)Bobi{\'c}, Lalonde, Nesteruk, Lee, Nenoff, Gorissen, Bertolet, Busse, Chan, Winey, et~al.]{bobic2023large}
Mislav Bobi{\'c}, Arthur Lalonde, Konrad~P Nesteruk, Hoyeon Lee, Lena Nenoff, Bram~L Gorissen, Alejandro Bertolet, Paul~M Busse, Annie~W Chan, Brian~A Winey, et~al.
\newblock Large anatomical changes in head-and-neck cancers--a dosimetric comparison of online and offline adaptive proton therapy.
\newblock \emph{Clinical and Translational Radiation Oncology}, 40:\penalty0 100625, 2023.

\bibitem[Burlacu et~al.(2025)Burlacu, Hoogeman, Lathouwers, and Perk{\'o}]{burlacu2025deep}
Tiberiu Burlacu, Mischa Hoogeman, Danny Lathouwers, and Zolt{\'a}n Perk{\'o}.
\newblock A deep learning model for inter-fraction head and neck anatomical changes in proton therapy.
\newblock \emph{Physics in Medicine and Biology}, 2025.

\bibitem[Castelli et~al.(2015)Castelli, Simon, Louvel, Henry, Chajon, Nassef, Haigron, Cazoulat, Ospina, Jegoux, et~al.]{castelli2015impact}
Joel Castelli, Antoine Simon, Guillaume Louvel, Olivier Henry, Enrique Chajon, Mohamed Nassef, Pascal Haigron, Guillaume Cazoulat, Juan~David Ospina, Franck Jegoux, et~al.
\newblock Impact of head and neck cancer adaptive radiotherapy to spare the parotid glands and decrease the risk of xerostomia.
\newblock \emph{Radiation Oncology}, 10:\penalty0 1--10, 2015.

\bibitem[Caudell et~al.(2017)Caudell, Torres-Roca, Gillies, Enderling, Kim, Rishi, Moros, and Harrison]{caudell2017future}
Jimmy~J Caudell, Javier~F Torres-Roca, Robert~J Gillies, Heiko Enderling, Sungjune Kim, Anupam Rishi, Eduardo~G Moros, and Louis~B Harrison.
\newblock The future of personalised radiotherapy for head and neck cancer.
\newblock \emph{The Lancet Oncology}, 18\penalty0 (5):\penalty0 e266--e273, 2017.

\bibitem[Chao et~al.(2001)Chao, Deasy, Markman, Haynie, Perez, Purdy, and Low]{chao2001prospective}
KS~Clifford Chao, Joseph~O Deasy, Jerry Markman, Joyce Haynie, Carlos~A Perez, James~A Purdy, and Daniel~A Low.
\newblock A prospective study of salivary function sparing in patients with head-and-neck cancers receiving intensity-modulated or three-dimensional radiation therapy: initial results.
\newblock \emph{International Journal of Radiation Oncology, Biology, Physics}, 49\penalty0 (4):\penalty0 907--916, 2001.

\bibitem[Ebrahimi and Lim(2021)]{ebrahimi2021reinforcement}
Saba Ebrahimi and Gino~J Lim.
\newblock A reinforcement learning approach for finding optimal policy of adaptive radiation therapy considering uncertain tumor biological response.
\newblock \emph{Artificial Intelligence in Medicine}, 121:\penalty0 102193, 2021.

\bibitem[Eckardt et~al.(2021)Eckardt, Wendt, Bornhaeuser, and Middeke]{eckardt2021reinforcement}
Jan-Niklas Eckardt, Karsten Wendt, Martin Bornhaeuser, and Jan~Moritz Middeke.
\newblock Reinforcement learning for precision oncology.
\newblock \emph{Cancers}, 13\penalty0 (18):\penalty0 4624, 2021.

\bibitem[Fu et~al.(2023)Fu, Taasti, and Zarepisheh]{fu2023distributed}
Anqi Fu, Vicki~T Taasti, and Masoud Zarepisheh.
\newblock Distributed and scalable optimization for robust proton treatment planning.
\newblock \emph{Medical Physics}, 50\penalty0 (1):\penalty0 633--642, 2023.

\bibitem[Gao et~al.(2024)Gao, Park, and Jia]{gao2024human}
Yin Gao, Yang~Kyun Park, and Xun Jia.
\newblock Human-like intelligent automatic treatment planning of head and neck cancer radiation therapy.
\newblock \emph{Physics in Medicine \& Biology}, 69\penalty0 (11):\penalty0 115049, 2024.

\bibitem[Ghobadi et~al.(2012)Ghobadi, Ghaffari, Aleman, Jaffray, and Ruschin]{ghobadi2012automated}
Kimia Ghobadi, Hamid~R Ghaffari, Dionne~M Aleman, David~A Jaffray, and Mark Ruschin.
\newblock Automated treatment planning for a dedicated multi-source intracranial radiosurgery treatment unit using projected gradient and grassfire algorithms.
\newblock \emph{Medical Physics}, 39\penalty0 (6(1)):\penalty0 3134--3141, 2012.

\bibitem[Gorissen(2022)]{gorissen2022interior}
Bram~L Gorissen.
\newblock Interior point methods can exploit structure of convex piecewise linear functions with application in radiation therapy.
\newblock \emph{SIAM Journal on Optimization}, 32\penalty0 (1):\penalty0 256--275, 2022.

\bibitem[Holliday et~al.(2015)Holliday, Garden, Rosenthal, Fuller, Morrison, Gunn, Phan, Beadle, Zhu, Zhang, et~al.]{holliday2015proton}
Emma~B Holliday, Adam~S Garden, David~I Rosenthal, C~David Fuller, William~H Morrison, G~Brandon Gunn, Jack Phan, Beth~M Beadle, Xiarong~R Zhu, Xiaodong Zhang, et~al.
\newblock Proton therapy reduces treatment-related toxicities for patients with nasopharyngeal cancer: a case-match control study of intensity-modulated proton therapy and intensity-modulated photon therapy.
\newblock \emph{International Journal of Particle Therapy}, 2\penalty0 (1):\penalty0 19--28, 2015.

\bibitem[Huiskes et~al.(2023)Huiskes, Astreinidou, Kong, Breedveld, Heijmen, and Rasch]{huiskes2023dosimetric}
Merle Huiskes, Eleftheria Astreinidou, Wens Kong, Sebastiaan Breedveld, Ben Heijmen, and Coen Rasch.
\newblock Dosimetric impact of adaptive proton therapy in head and neck cancer--a review.
\newblock \emph{Clinical and Translational Radiation Oncology}, 39:\penalty0 100598, 2023.

\bibitem[Kim et~al.(2018)Kim, Leeman, Riaz, McBride, Tsai, and Lee]{kim2018proton}
Joseph~K Kim, Jonathan~E Leeman, Nadeem Riaz, Sean McBride, Chiaojung~Jillian Tsai, and Nancy~Y Lee.
\newblock Proton therapy for head and neck cancer.
\newblock \emph{Current Treatment Options in Oncology}, 19:\penalty0 1--14, 2018.

\bibitem[Leeman et~al.(2017)Leeman, Romesser, Zhou, McBride, Riaz, Sherman, Cohen, Cahlon, and Lee]{leeman2017proton}
Jonathan~E Leeman, Paul~B Romesser, Ying Zhou, Sean McBride, Nadeem Riaz, Eric Sherman, Marc~A Cohen, Oren Cahlon, and Nancy Lee.
\newblock Proton therapy for head and neck cancer: expanding the therapeutic window.
\newblock \emph{The Lancet Oncology}, 18\penalty0 (5):\penalty0 e254--e265, 2017.

\bibitem[Li et~al.(2020)Li, Lee, Cohen, Sherman, and Lee]{li2020past}
Xingzhe Li, Anna Lee, Marc~A Cohen, Eric~J Sherman, and Nancy~Y Lee.
\newblock Past, present and future of proton therapy for head and neck cancer.
\newblock \emph{Oral Oncology}, 110:\penalty0 104879, 2020.

\bibitem[Ma et~al.(2022)Ma, Chen, Wang, Qin, Yan, Cao, Chen, Dai, and Men]{ma2022personalized}
Xiangyu Ma, Xinyuan Chen, Yu~Wang, Shirui Qin, Xuena Yan, Ying Cao, Yan Chen, Jianrong Dai, and Kuo Men.
\newblock Personalized modeling to improve pseudo--computed tomography images for magnetic resonance imaging--guided adaptive radiation therapy.
\newblock \emph{International Journal of Radiation Oncology, Biology, Physics}, 113\penalty0 (4):\penalty0 885--892, 2022.

\bibitem[Maniscalco et~al.(2023)Maniscalco, Liang, Lin, Jiang, and Nguyen]{maniscalco2023intentional}
Austen Maniscalco, Xiao Liang, Mu-Han Lin, Steve Jiang, and Dan Nguyen.
\newblock Intentional deep overfit learning for patient-specific dose predictions in adaptive radiotherapy.
\newblock \emph{Medical Physics}, 50\penalty0 (9):\penalty0 5354--5363, 2023.

\bibitem[McKeever et~al.(2016)McKeever, Sio, Gunn, Holliday, Blanchard, Kies, Weber, and Frank]{mckeever2016reduced}
Matthew~R McKeever, Terence~T Sio, G~Brandon Gunn, Emma~B Holliday, Pierre Blanchard, Merrill~S Kies, Randal~S Weber, and Steven~J Frank.
\newblock Reduced acute toxicity and improved efficacy from intensity-modulated proton therapy (impt) for the management of head and neck cancer.
\newblock \emph{Chinese Clinical Oncology}, 5\penalty0 (4):\penalty0 54--54, 2016.

\bibitem[Mnih et~al.(2015)Mnih, Kavukcuoglu, Silver, Rusu, Veness, Bellemare, Graves, Riedmiller, Fidjeland, Ostrovski, et~al.]{mnih2015human}
Volodymyr Mnih, Koray Kavukcuoglu, David Silver, Andrei~A Rusu, Joel Veness, Marc~G Bellemare, Alex Graves, Martin Riedmiller, Andreas~K Fidjeland, Georg Ostrovski, et~al.
\newblock Human-level control through deep reinforcement learning.
\newblock \emph{Nature}, 518\penalty0 (7540):\penalty0 529--533, 2015.

\bibitem[Moreau et~al.(2021)Moreau, Fran{\c{c}}ois-Lavet, Desbordes, and Macq]{moreau2021reinforcement}
Gr{\'e}goire Moreau, Vincent Fran{\c{c}}ois-Lavet, Paul Desbordes, and Beno{\^\i}t Macq.
\newblock Reinforcement learning for radiotherapy dose fractioning automation.
\newblock \emph{Biomedicines}, 9\penalty0 (2):\penalty0 214, 2021.

\bibitem[Moreno et~al.(2019)Moreno, Frank, Garden, Rosenthal, Fuller, Gunn, Reddy, Morrison, Williamson, Holliday, et~al.]{moreno2019intensity}
Amy~C Moreno, Steven~J Frank, Adam~S Garden, David~I Rosenthal, Clifton~D Fuller, Gary~B Gunn, Jay~P Reddy, William~H Morrison, Tyler~D Williamson, Emma~B Holliday, et~al.
\newblock Intensity modulated proton therapy (impt)--the future of imrt for head and neck cancer.
\newblock \emph{Oral Oncology}, 88:\penalty0 66--74, 2019.

\bibitem[Nelms et~al.(2012)Nelms, Robinson, Markham, Velasco, Boyd, Narayan, Wheeler, and Sobczak]{nelms2012variation}
Benjamin~E Nelms, Greg Robinson, Jay Markham, Kyle Velasco, Steve Boyd, Sharath Narayan, James Wheeler, and Mark~L Sobczak.
\newblock Variation in external beam treatment plan quality: an inter-institutional study of planners and planning systems.
\newblock \emph{Practical Radiation Oncology}, 2\penalty0 (4):\penalty0 296--305, 2012.

\bibitem[Nikou et~al.(2024)Nikou, Thompson, Nisbet, Gulliford, and McClelland]{nikou2024modelling}
Poppy Nikou, Anna Thompson, Andrew Nisbet, Sarah Gulliford, and Jamie McClelland.
\newblock Modelling systematic anatomical uncertainties of head and neck cancer patients during fractionated radiotherapy treatment.
\newblock \emph{Physics in Medicine \& Biology}, 69\penalty0 (15):\penalty0 155017, 2024.

\bibitem[Nocedal and Wright(1999)]{nocedal1999numerical}
Jorge Nocedal and Stephen~J Wright.
\newblock \emph{Numerical optimization}.
\newblock Springer, 1999.

\bibitem[Schulman et~al.(2016)Schulman, Moritz, Levine, Jordan, and Abbeel]{schulman2015high}
John Schulman, Philipp Moritz, Sergey Levine, Michael Jordan, and Pieter Abbeel.
\newblock High-dimensional continuous control using generalized advantage estimation.
\newblock \emph{Proceedings of the International Conference on Learning Representations (ICLR)}, 2016.

\bibitem[Schulman et~al.(2017)Schulman, Wolski, Dhariwal, Radford, and Klimov]{schulman2017proximal}
John Schulman, Filip Wolski, Prafulla Dhariwal, Alec Radford, and Oleg Klimov.
\newblock Proximal policy optimization algorithms.
\newblock \emph{arXiv preprint arXiv:1707.06347}, 2017.

\bibitem[Shen et~al.(2019)Shen, Gonzalez, Klages, Qin, Jung, Chen, Nguyen, Jiang, and Jia]{shen2019intelligent}
Chenyang Shen, Yesenia Gonzalez, Peter Klages, Nan Qin, Hyunuk Jung, Liyuan Chen, Dan Nguyen, Steve~B Jiang, and Xun Jia.
\newblock Intelligent inverse treatment planning via deep reinforcement learning, a proof-of-principle study in high dose-rate brachytherapy for cervical cancer.
\newblock \emph{Physics in Medicine \& Biology}, 64\penalty0 (11):\penalty0 115013, 2019.

\bibitem[Shen et~al.(2020)Shen, Nguyen, Chen, Gonzalez, McBeth, Qin, Jiang, and Jia]{shen2020operating}
Chenyang Shen, Dan Nguyen, Liyuan Chen, Yesenia Gonzalez, Rafe McBeth, Nan Qin, Steve~B Jiang, and Xun Jia.
\newblock Operating a treatment planning system using a deep-reinforcement learning-based virtual treatment planner for prostate cancer intensity-modulated radiation therapy treatment planning.
\newblock \emph{Medical Physics}, 47\penalty0 (6):\penalty0 2329--2336, 2020.

\bibitem[Shen et~al.(2021{\natexlab{a}})Shen, Chen, Gonzalez, and Jia]{shen2021improving}
Chenyang Shen, Liyuan Chen, Yesenia Gonzalez, and Xun Jia.
\newblock Improving efficiency of training a virtual treatment planner network via knowledge-guided deep reinforcement learning for intelligent automatic treatment planning of radiotherapy.
\newblock \emph{Medical Physics}, 48\penalty0 (4):\penalty0 1909--1920, 2021{\natexlab{a}}.

\bibitem[Shen et~al.(2021{\natexlab{b}})Shen, Chen, and Jia]{shen2021hierarchical}
Chenyang Shen, Liyuan Chen, and Xun Jia.
\newblock A hierarchical deep reinforcement learning framework for intelligent automatic treatment planning of prostate cancer intensity modulated radiation therapy.
\newblock \emph{Physics in Medicine \& Biology}, 66\penalty0 (13):\penalty0 134002, 2021{\natexlab{b}}.

\bibitem[Sonke et~al.(2019)Sonke, Aznar, and Rasch]{sonke2019adaptive}
Jan-Jakob Sonke, Marianne~C. Aznar, and Coen R.~N. Rasch.
\newblock Adaptive radiotherapy for anatomical changes.
\newblock \emph{Seminars in Radiation Oncology}, 29 3:\penalty0 245--257, 2019.

\bibitem[Sutton et~al.(1998)Sutton, Barto, et~al.]{sutton1998reinforcement}
Richard~S Sutton, Andrew~G Barto, et~al.
\newblock \emph{Reinforcement learning: An Introduction}, volume~1.
\newblock MIT Press Cambridge, 1998.

\bibitem[Visak et~al.(2024)Visak, Liao, Zhong, Wang, Domal, Wang, Maniscalco, Pompos, Nyguen, Parsons, et~al.]{visak2024assessing}
Justin Visak, Chien-Yi Liao, Xinran Zhong, Biling Wang, Sean Domal, Hui-Ju Wang, Austen Maniscalco, Arnold Pompos, Dan Nyguen, David Parsons, et~al.
\newblock Assessing population-based to personalized planning strategies for head and neck adaptive radiotherapy.
\newblock \emph{Journal of Applied Clinical Medical Physics}, page e14576, 2024.

\bibitem[Volpe et~al.(2021)Volpe, Pepa, Zaffaroni, Bellerba, Santamaria, Marvaso, Isaksson, Gandini, Starzy{\'n}ska, Leonardi, et~al.]{volpe2021machine}
Stefania Volpe, Matteo Pepa, Mattia Zaffaroni, Federica Bellerba, Riccardo Santamaria, Giulia Marvaso, Lars~Johannes Isaksson, Sara Gandini, Anna Starzy{\'n}ska, Maria~Cristina Leonardi, et~al.
\newblock Machine learning for head and neck cancer: a safe bet?—a clinically oriented systematic review for the radiation oncologist.
\newblock \emph{Frontiers in Oncology}, 11:\penalty0 772663, 2021.

\bibitem[Wang and Chang(2024)]{wang2024automating}
Qingqing Wang and Chang Chang.
\newblock Automating proton pbs treatment planning for head and neck cancers using policy gradient-based deep reinforcement learning.
\newblock \emph{arXiv preprint arXiv:2409.11576}, 2024.

\bibitem[Watkins and Dayan(1992)]{watkins1992q}
Christopher~JCH Watkins and Peter Dayan.
\newblock Q-learning.
\newblock \emph{Machine Learning}, 8:\penalty0 279--292, 1992.

\bibitem[Wieser et~al.(2017)Wieser, Cisternas, Wahl, Ulrich, Stadler, Mescher, M{\"u}ller, Klinge, Gabrys, Burigo, et~al.]{wieser2017development}
Hans-Peter Wieser, Eduardo Cisternas, Niklas Wahl, Silke Ulrich, Alexander Stadler, Henning Mescher, Lucas-Raphael M{\"u}ller, Thomas Klinge, Hubert Gabrys, Lucas Burigo, et~al.
\newblock Development of the open-source dose calculation and optimization toolkit matrad.
\newblock \emph{Medical Physics}, 44\penalty0 (6):\penalty0 2556--2568, 2017.

\bibitem[Wildman et~al.(2024)Wildman, Wynne, Kesarwala, and Yang]{wildman2024recent}
Vanessa~L Wildman, Jacob~F Wynne, Aparna~H Kesarwala, and Xiaofeng Yang.
\newblock Recent advances in the clinical applications of machine learning in proton therapy.
\newblock \emph{medRxiv}, pages 2024--10, 2024.

\bibitem[Yang et~al.(2024)Yang, Wu, Li, Mansfield, Xie, Wu, Wu, and Sheng]{yang2024automated}
Dongrong Yang, Xin Wu, Xinyi Li, Ryan Mansfield, Yibo Xie, Qiuwen Wu, Q~Jackie Wu, and Yang Sheng.
\newblock Automated treatment planning with deep reinforcement learning for head-and-neck (hn) cancer intensity modulated radiation therapy (imrt).
\newblock \emph{Physics in Medicine \& Biology}, 70\penalty0 (1):\penalty0 015010, 2024.

\bibitem[Zhang et~al.(2020)Zhang, Wang, Sheng, Palta, Czito, Willett, Zhang, Jensen, Yin, Wu, Ge, and Wu]{Zhang2020AnIP}
Jiahan Zhang, C.~Wang, Yang Sheng, Manisha Palta, Brian~G Czito, Christopher~G Willett, Jiang Zhang, P~James Jensen, Fang-Fang Yin, Qiuwen Wu, Yaorong Ge, and Q.~Jackie Wu.
\newblock An interpretable planning bot for pancreas stereotactic body radiation therapy.
\newblock \emph{International Journal of Radiation Oncology, Biology, Physics}, 2020.

\end{thebibliography}

\newpage
\appendix

\section{Algorithm: Patient-Specific DRL for IMPT Replanning}
\label{appndx:alg_protonRL}

\begin{algorithm2e}[H]
    \caption{Patient-Specific DRL for IMPT Replanning}
    \label{alg:drl_impt_replan}
    \small
    \LinesNumbered
    \DontPrintSemicolon
        \KwIn{Patient cohort $\mathcal{P}$, RL algorithm (DQN/PPO), total episodes $N_{\text{episodes}}$, planning horizon $T_{\text{max}}$, learning rate $\eta$, target update rate $\tau$}
        \KwOut{Personalized policies $\{\pi_p\}_{p \in \mathcal{P}}$}
        \ForEach{patient $p \in \mathcal{P}$}{
            \textbf{Initial Plan Setup:}\;
            Load dose matrices $\Vec{D}^p$, prescriptions (Rx), and initial priorities $\{\Vec{\omega}^p\}$\;
            Initialize beamlet intensity vector $\Vec{x}_0 \leftarrow \Vec{1}$\;
            Compute initial dose distribution $\Vec{d}_0 \leftarrow \Vec{D}^p\Vec{x}_0$\;
            Generate DVH curves as initial state $s_0^p \leftarrow \text{StateGen}(\Vec{d}_0)$\;
            Compute initial plan quality score (0-150 scale): $R_{\text{prev}}^p \leftarrow \psi(s_0^p)$\;
            
            \textbf{RL Training Setup:}\;
            Initialize policy network $\pi_{\Vec{\theta}_p}$ with weights $\Vec{\theta}_p^0$\;
            \If{DQN}{
                Initialize target network $\Vec{\theta}_p^- \leftarrow \Vec{\theta}_p^0$\;
                Create replay buffer $\mathcal{B}_p \leftarrow \emptyset$\;
            }
            
            \textbf{Training Loop:}\;
            \For{episode $= 1$ \KwTo $N_{\text{episodes}}$}{
                Reset environment: $s_t^p \leftarrow s_0^p$, $\Vec{x}_t \leftarrow \Vec{x}_0$\;
                \For{step $t = 0$ \KwTo $T_{\text{max}}$}{
                    Select priority adjustment action $a_t^p \sim \pi_{\Vec{\theta}_p}(s_t^p)$\;
                    Adjust tuning priorities: $\Vec{\omega}_{t+1}^p \leftarrow \text{UpdateTPP}(\Vec{\omega}_t^p, a_t^p)$ (Eqn.~\ref{eq:priority_update_rule})\;
                    Optimize beamlet intensities: $\Vec{x}_{t+1} \leftarrow \arg\min_{\Vec{x} \geq 0} \mathcal{L}$ (Eqn.~\ref{eq:plan_optimization_problem})\;
                    Compute new dose $\Vec{d}_{t+1} \leftarrow \Vec{D}^p\Vec{x}_{t+1}$ (Eqn.~\ref{eq:dij_mat})\;
                    Extract state: $s_{t+1}^p \leftarrow \text{StateGen}(\Vec{d}_{t+1})$\;
                    Calculate plan quality score $R_{t+1}^p \leftarrow \psi(s_{t+1}^p)$\;
                    Calculate reward $r_{t+1}^p \leftarrow R_{t+1}^p - R_{\text{prev}}^p$\;
                    \If{DQN}{
                        Store transition $(s_t^p, a_t^p, r_{t+1}^p, s_{t+1}^p)$ in $\mathcal{B}_p$\;
                        Sample batch $b \sim \mathcal{B}_p$, update $\Vec{\theta}_p \leftarrow \Vec{\theta}_p - \eta\nabla\mathcal{L}_{\text{DQN}}(b)$ (Eqn.~\ref{eq:dqn_loss})\;
                        Soft update: $\Vec{\theta}_p^- \leftarrow \tau\Vec{\theta}_p + (1-\tau)\Vec{\theta}_p^-$\;
                    }
                    \If{PPO}{
                        Estimate advantage $\hat{A}_t(s_t^p, a_t^p)$ (Eqn.~\ref{eq:gae}) \; 
                        Update $\Vec{\theta}_p \leftarrow \Vec{\theta}_p - \eta\nabla\mathcal{L}_{\text{PPO}}(\hat{A}_t(s_t^p, a_t^p))$ (Eqn.~\ref{eq:ppo_loss})\;
                    }
                    \If{$R_{t+1}^p = 150.00$}{
                        Break 
                    }
                    Update next state $s_t^p \leftarrow s_{t+1}^p$\;
                    Update previous plan score $R_{\text{prev}}^p \leftarrow R_{t+1}^p$\;
                }
            }
        }
\end{algorithm2e}

\section{Cohort Demographics and Staging}
\label{appndx:cohort}

Table~\ref{tab:patient_demographics} summarizes the demographic and clinical staging data for the eight patients included in this study. Patient ages range from $48$ to $73$ years. The cohort encompasses a range of disease stages, including stage II ($n=2$), stage III ($n=3$), stage IVa ($n=1$), stage IVb ($n=1$), and one unassigned case. Detailed TNM classifications are also reported, reflecting the heterogeneity of tumor and nodal involvement across the sample.

\begin{table}[ht]
    \centering
    \caption{\textbf{Patient Demographics and Staging Information}.}
    \label{tab:patient_demographics}
        \begin{tabular}{cccc}
            \toprule
            \textbf{Case ID} & \textbf{Age} & \textbf{Overall Stage} & \textbf{TNM Stage} \\
            \midrule
            P1 & 69 & IVb & T0 N3 M0 \\
            P2 & 73 & Unassigned & T4 \& T2 N2 M0 \\
            P3 & 60 & II & T3 N2 M0 \\
            P4 & 55 & III & T4 N0 M0 \\
            P5 & 48 & II & T2 N2 M0 \\
            P6 & 72 & IVa & T4 N0 M0 \\
            P7 & 71 & III & T4 N0 M0 \\
            P8 & 59 & III & T4 N1 M0 \\
            \bottomrule
        \end{tabular}
\end{table}

\section{Priority Adjustments: Structure-Specific Modifications}
\label{appndx:priority_updates}

All replanning agents modify the treatment plan optimization priorities through $22$ discrete adjustments to weight parameters. The complete mapping is detailed in Table~\ref{tab:priority_adjustments}, organized by structure type and adjustment magnitude. This discrete action space enables clinically meaningful trade-off adjustments while maintaining valid priority ranges. For critical OARs (brainstem, spinal cord), larger adjustments ($+0.3$) are available to prioritize their protection, while other structures have more moderate adjustments. Target volumes include both positive and negative adjustments to allow for both improved coverage and compromises when necessary for OAR sparing.
\begin{table}[ht]
    \centering
    \caption{\textbf{Priority Adjustment Action Space.} Each action modifies the weight of a specific structure by a predefined increment.}
    \label{tab:priority_adjustments}
    \begin{tabular}{clcc}
        \toprule
        \textbf{Action Index} & \textbf{Structure} & \textbf{Parameter} & \textbf{Adjustment ($\Delta_a$)} \\
        \midrule
        \multicolumn{4}{l}{\textit{Target Volume Adjustments}} \\
        0, 1 & CTV1 & $\omega_\text{\tiny CTV1}$ & $+0.2, -0.1$ \\
        2, 3 & CTV2 & $\omega_\text{\tiny CTV2}$ & $+0.2, -0.1$ \\
        4, 5 & CTV3 & $\omega_\text{\tiny CTV3}$ & $+0.2, -0.1$ \\
        18, 19 & CTV2 & $\omega_a$ & $+0.1, -0.1$ \\
        20, 21 & CTV3 & $\omega_b$ & $+0.1, -0.1$ \\
        \midrule
        \multicolumn{4}{l}{\textit{Organ-at-Risk Adjustments}} \\
        6 & Mandible (MAN) & $\omega_\text{\tiny MAN}$ & $+0.1$ \\
        7 & Brainstem (BRS) & $\omega_\text{\tiny BRS}$ & $+0.3$ \\
        8 & Spinal Cord (SC) & $\omega_\text{\tiny SC}$ & $+0.3$ \\
        9, 10 & Parotids (PARR, PARL) & $\omega_\text{\tiny PAR(L/R)}$ & $+0.1$ \\
        11 & Larynx (LAR) & $\omega_\text{\tiny LAR}$ & $+0.1$ \\
        12 & Pharynx (PHY) & $\omega_\text{\tiny PHY}$ & $+0.1$ \\
        13, 14 & Cochleas (COCHL, COCHR) & $\omega_\text{\tiny COCH(L/R)}$ & $+0.1$ \\
        15, 16 & Submand. Glands (SMGL, SMGR) & $\omega_\text{\tiny SMG(L/R)}$ & $+0.1$ \\
        17 & Esophagus (ESO) & $\omega_\text{\tiny ESO}$ & $+0.1$ \\
        \bottomrule
    \end{tabular}
\end{table}

\section{Plan Quality Metrics}
\label{appndx:pqm}

\begin{figure}[t]
   \centering 
   \includegraphics[width=\textwidth]{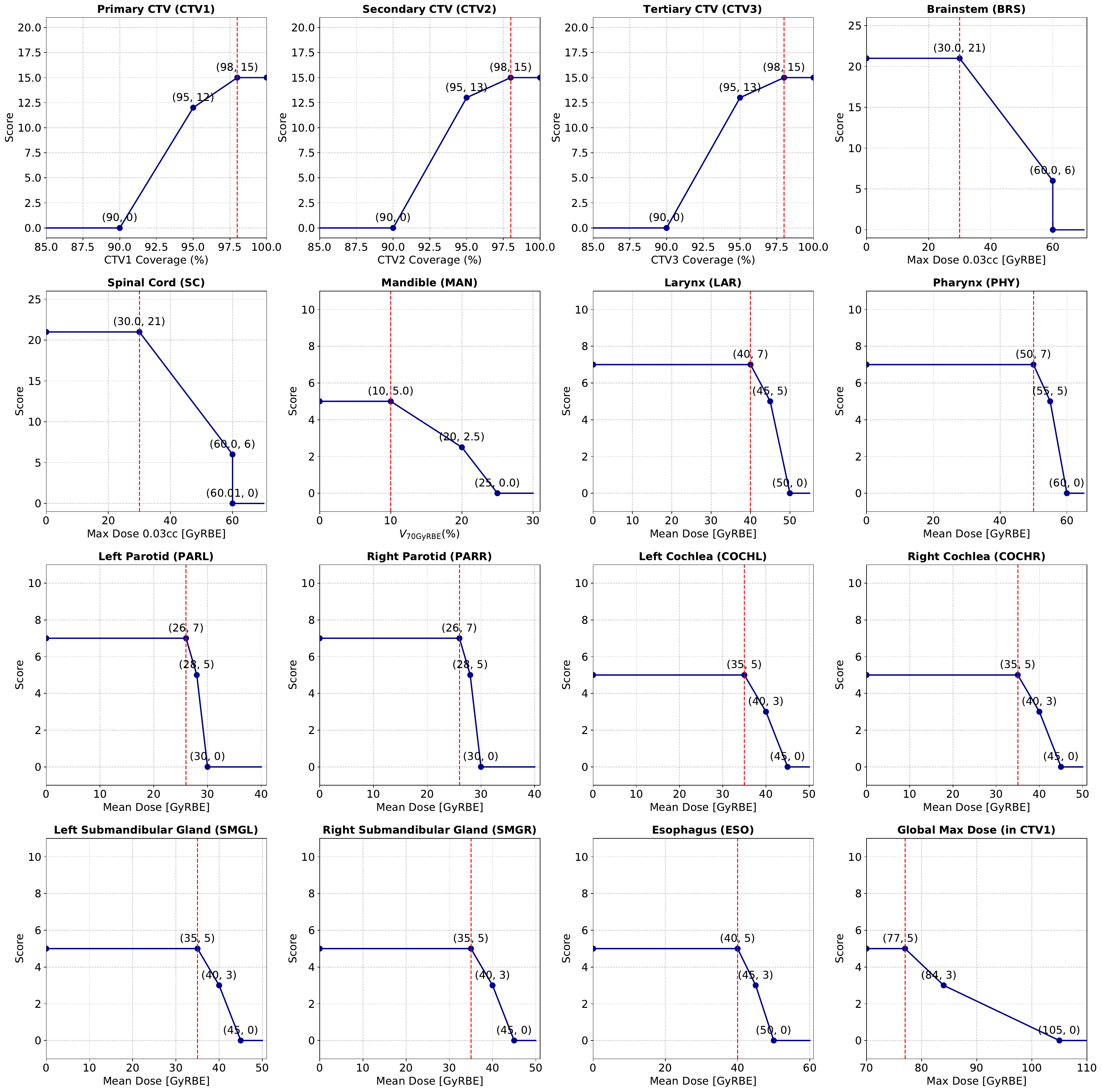} 
   \caption{\textbf{Scoring functions for different dosimetric parameters.} ProKnow-based scoring functions for quantifying the quality of treatment plans. The vertical red-dashed lines denote clinical planning goals. Plan quality is the sum of all components (max score = 150.0).}
   \label{fig:proknow_metrics} 
 \end{figure}

Plan quality was quantified using a comprehensive 150-point scoring system derived from standardized ProKnow criteria~\citep{nelms2012variation} and institutional planning guidelines. This total score is the sum of individual component scores (detailed in Figure~\ref{fig:proknow_metrics}), which directly relate to the planning objectives outlined in Table~\ref{tab:planning_constraints}. For reinforcement learning, the reward signal was defined as the change in this cumulative plan quality score between tuning steps.

\section{Network Architectures \& Training Parameters}
\label{appndx:network_details}

Both DRL agents process the DVHs using Convolutional Neural Networks (CNNs). This architectural choice is motivated by the ability of CNNs to learn meaningful spatial features and correlations from the structured DVH input, capturing complex dose-volume relationships across different anatomical structures. Common elements in both architectures include ReLU activation functions, batch normalization layers following convolutions, max-pooling for dimensionality reduction, and dropout layers for regularization. The DQN agent employs two convolutional layers followed by three fully connected layers to estimate Q-values for $22$ discrete actions. The DQN implementation features experience replay with prioritized sampling based on temporal difference errors, exponential epsilon-greedy decay, and a delayed target network updated every $10$ steps to stabilize learning. The PPO agent adopts a shared actor-critic architecture with a three-layer CNN backbone. The actor head outputs $22$ discrete priority adjustments via a categorical distribution, while the critic estimates state values. Training incorporates clipped policy updates ($\epsilon=0.2$), generalized advantage estimation (GAE, $\lambda=0.95$), value loss coefficient ($c_1=1.00$), and entropy coefficient ($c_2=0.01$) to balance exploration and exploitation. Orthogonal initialization was applied to weights in the PPO network, with constant initialization for biases. Both algorithms use the Adam optimizer with learning rate $1e-5$, with PPO additionally employing gradient clipping (max norm $=0.5$) to ensure training stability. Table~\ref{tab:architectures} details the architectural specifications for both models. Table~\ref{tab:training_params} highlights the key hyperparameters and optimization settings used in training.

\begin{table}[ht]
    \centering
    \caption{\textbf{Network architectures for DQN and PPO agents.}}
    \label{tab:architectures}
        \begin{tabular}{lcc}
            \toprule
            \textbf{Component} & \textbf{DQN} & \textbf{PPO} \\
            \midrule
            Input Dimension & \multicolumn{2}{c}{15 × 100 (Structures × Dose Bins)} \\
            Input Channels & 1 & 1 \\
            Convolutional Layers & 2 & 3 \\
            Kernel Sizes & (3,5) & (3,5) \\
            Padding & (1,2) & (1,2) \\
            Pooling & 2×2 MaxPool & 2×2 MaxPool \\
            BatchNorm & Yes & Yes \\
            Dropout Rate & 0.2 & 0.2 \\
            Hidden Dim (CNN) & 32 → 64 & 32 → 64 → 64 \\
            Fully Connected & 256 → 128 → 22 & 512 → 22 (Actor) \\
            Output Activation & Linear & Softmax (Actor) \\
            \bottomrule
        \end{tabular}
\end{table}

\begin{table}[ht]
    \centering
    \caption{\textbf{Training Parameters for DQN and PPO Agents.}}
    \label{tab:training_params}
        \begin{tabular}{lcc}
            \toprule
            \textbf{Parameter} & \textbf{DQN} & \textbf{PPO} \\
            \midrule
            Optimizer & Adam & Adam \\
            Learning Rate & 1e-5 & 1e-5 \\
            Minibatch Size & $16$ & $16$ \\
            Discount Factor ($\gamma$) & $0.99$ & $0.99$\\
            \midrule
            Loss Function & MSE & \makecell{Clipped Surrogate + Value \\ MSE + Entropy Bonus} \\
            & (Eqn.~\ref{eq:dqn_loss}) & (Eqn.~\ref{eq:ppo_loss}) \\
            Experience Buffer & Prioritized Replay & On-policy Rollouts \\
            Buffer Retention & Long-term & Episodic \\
            Epsilon-Greedy Exploration & Yes (exponential decay) & No \\
            Advantage Estimation  & -- & GAE ($\lambda = 0.95$) \\
            Clipped Policy Update & -- & $0.2 $\\
            Value Clip & -- & $0.2$ \\
            Value Loss Coefficient ($c_1$) & -- & $1.0$ \\
            Entropy Coefficient ($c_2$) & -- & $0.01$ \\
            Gradient Clipping & \textit{Optional} & $0.5$ (max norm) \\
            \bottomrule
        \end{tabular}
\end{table}

\section{Complete Dosimetric Results}
\label{appndx:complete_results}

This section contains the complete patient-specific results on the \nth{1} replanning CT, including DVH comparisons and detailed dosimetric metrics for the manual and DRL-based plans.

\begin{figure}[ht]
    \centering 
    \includegraphics[width=0.9\linewidth]{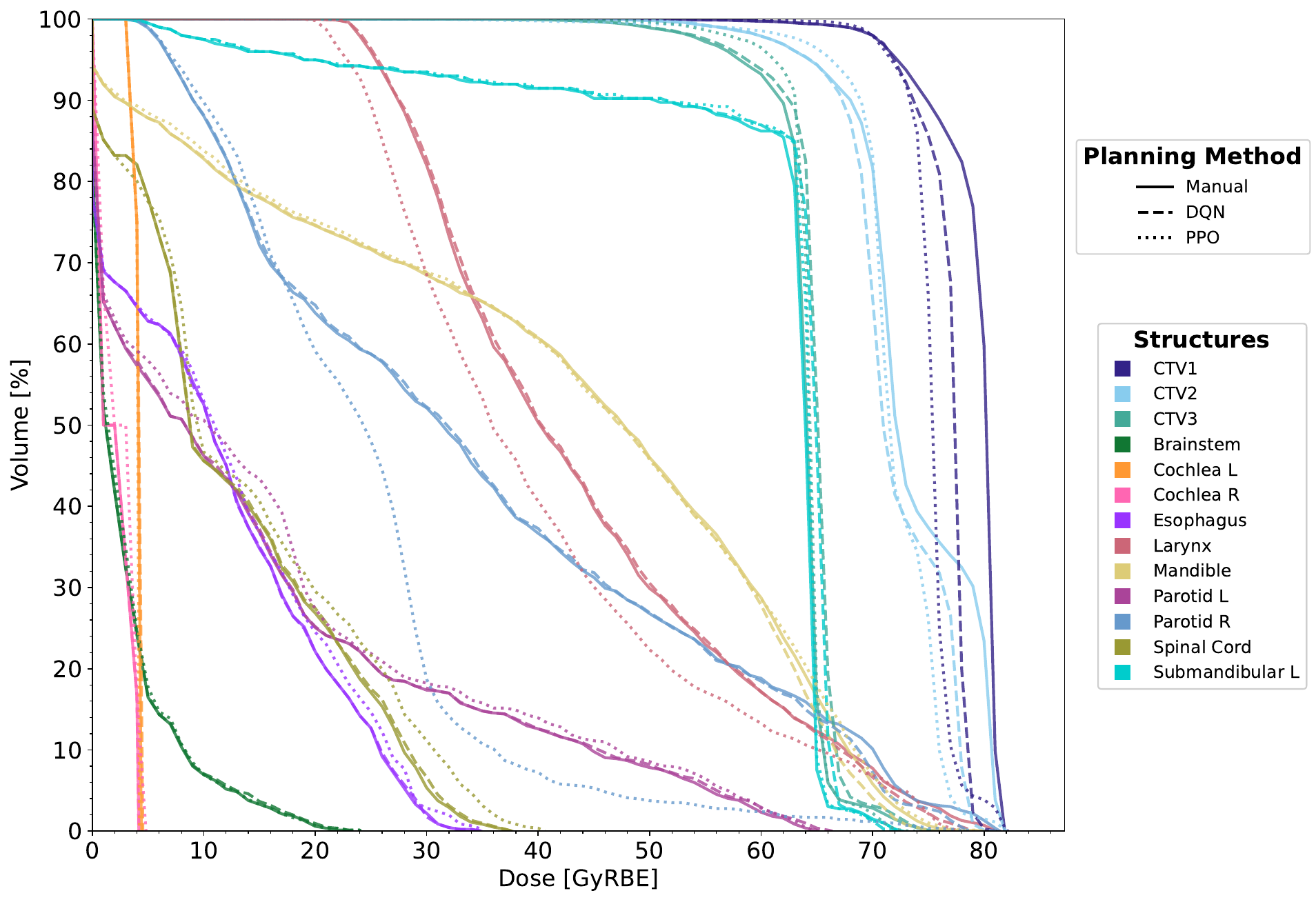} 
    \caption{\textbf{DVH comparison for patient P1's \nth{1} rpCT}: Manual replans (solid lines), patient-specific DQN (dashed lines), and patient-specific PPO (dotted lines).}
    \label{fig:P1_dvh} 
\end{figure}

\begin{table}[ht]
    \centering
    \caption{\textbf{Comparison of dosimetric endpoints across planning strategies for patient P1 on \nth{1} rpCT.} All dose values ($D_{\text{0\%/0.03\si{cc}/mean}}$) are in~\si{\rbe}.}
    \label{tab:P1_metrics}
    \small
    \resizebox{0.9\textwidth}{!}{%
        \begin{tabular}{lllc cc cc cc}
            \toprule
            \textbf{Delineated} & \multicolumn{2}{c}{\textbf{Dosimetric Endpoint}} & \multicolumn{2}{c}{\textbf{Manual}} & \multicolumn{2}{c}{\textbf{DQN}} & \multicolumn{2}{c}{\textbf{PPO}} \\
            \cmidrule(lr){2-3} \cmidrule(lr){4-5} \cmidrule(lr){6-7} \cmidrule(lr){8-9}
            \textbf{Structure} & \textbf{Metric} & \textbf{Score} & \textbf{Value} & \textbf{Score} & \textbf{Value} & \textbf{Score} & \textbf{Value} & \textbf{Score} \\
            \midrule
            \multirow{2}{*}{CTV1} 
            & $V_{d_{Rx,\text{CTV}_1}} \geq 98\%$ & 15 & 97.99 & 14.99 & 98.01 & 15.00 & 97.99 & 14.99 \\
            & $D_{max} \leq 77$ & 5 & 81.86 & 3.31 & 81.02 & 3.43 & 82.24 & 3.25 \\
            \midrule
            CTV2 & $V_{d_{Rx,\text{CTV}_2}} \geq 98\%$ & 15 & 97.97 & 14.98 & 97.99 & 14.99 & 98.56 & 15.00 \\
            CTV3 & $V_{d_{Rx,\text{CTV}_3}} \geq 98\%$ & 15 & 97.84 & 14.89 & 98.03 & 15.00 & 99.00 & 15.00 \\
            \midrule
            BRS & $D_{\text{0.03cc}} \leq 30$ & 21 & 23.29 & 21.00 & 24.01 & 21.00 & 24.17 & 21.00 \\
            SC & $D_{\text{0.03cc}} \leq 30$ & 21 & 37.63 & 17.18 & 37.67 & 17.17 & 40.60 & 15.70 \\
            MAN & $V_{70\si{\rbe}}  \leq 10\%$ & 5 & 1.73 & 5.00 & 0.58 & 5.00 & 2.18 & 5.00 \\
            LAR & $D_{\text{mean}} \leq 40$ & 7 & 43.76 & 5.50 & 43.84 & 5.46 & 40.15 & 6.94 \\
            PHY & $D_{\text{mean}} \leq 50$ & 7 & 0.00 & 7.00 & 0.00 & 7.00 & 0.00 & 7.00 \\
            PARL & $D_{\text{mean}} \leq 26$ & 7 & 14.43 & 7.00 & 14.57 & 7.00 & 15.56 & 7.00 \\
            PARR & $D_{\text{mean}} \leq 26$ & 7 & 34.60 & 0.00 & 34.50 & 0.00 & 23.86 & 7.00 \\
            COCHL & $D_{\text{mean}} \leq 35$ & 5 & 4.03 & 5.00 & 4.15 & 5.00 & 3.96 & 5.00 \\
            COCHR & $D_{\text{mean}} \leq 35$ & 5 & 2.09 & 5.00 & 2.07 & 5.00 & 2.42 & 5.00 \\
            SMGL & $D_{\text{mean}} \leq 35$ & 5 & 59.73 & 0 & 60.59 & 0 & 60.07 & 0 \\
            SMGR & $D_{\text{mean}} \leq 35$ & 5 & 0.00 & 5.00 & 0.00 & 5.00 & 0.00 & 5.00 \\
            ESO & $D_{\text{mean}} \leq 40$ & 5 & 11.09 & 5.00 & 11.09 & 5.00 & 11.57 & 5.00 \\
            \midrule
            \multicolumn{2}{l}{\textbf{Cumulative Total}} & \textbf{150} & \textbf{--} & 130.85 & \textbf{--} & 131.05 & \textbf{--} & \textbf{137.88} \\ 
            \bottomrule
        \end{tabular}
    }
\end{table}

\begin{figure}[ht]
    \centering 
    \includegraphics[width=0.9\linewidth]{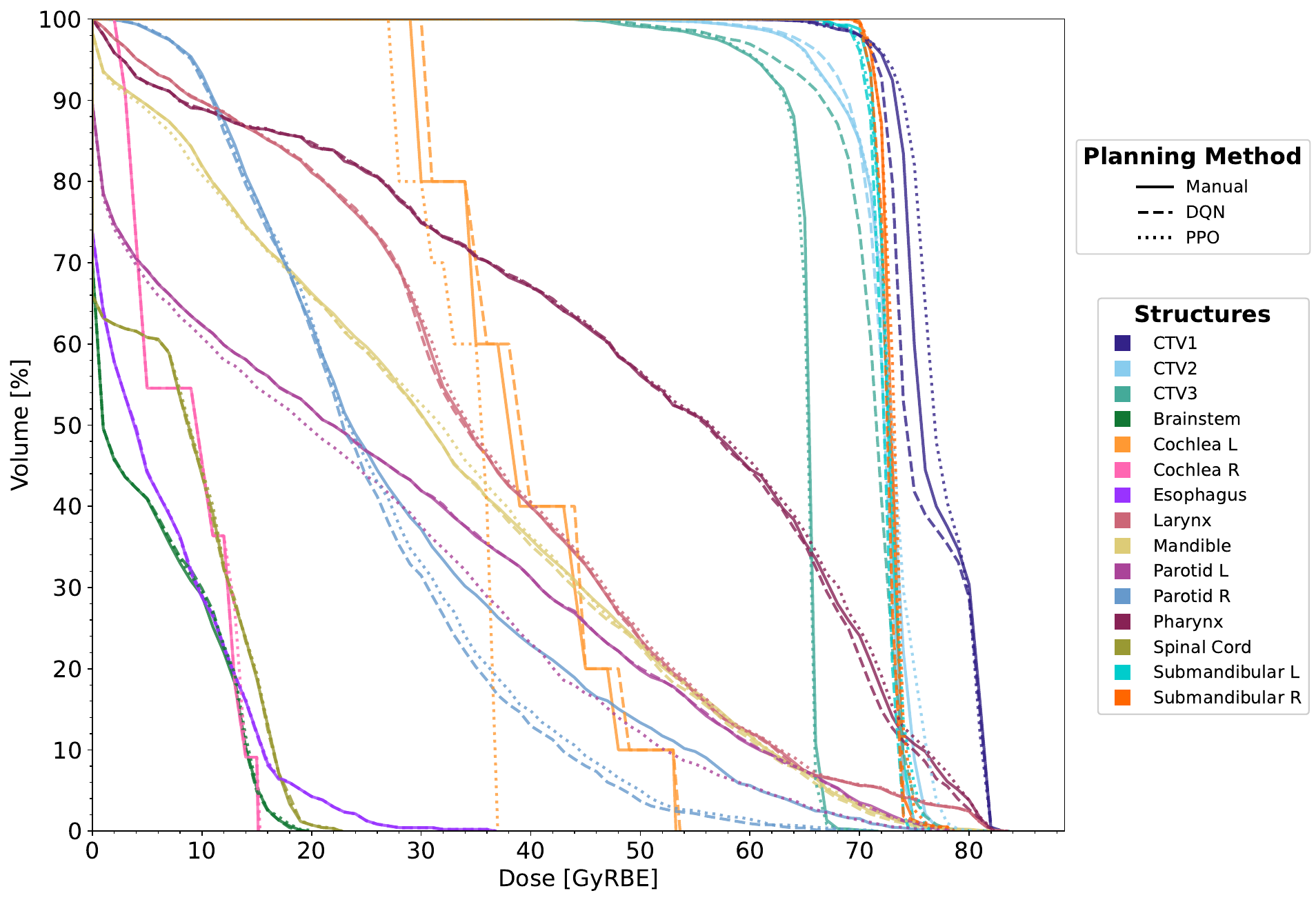} 
    \caption{\textbf{DVH comparison for patient P2's \nth{1} rpCT:} Manual replans (solid lines), patient-specific DQN (dashed lines), and patient-specific PPO (dotted lines).}
    \label{fig:P1_dvh_appndx} 
\end{figure}
 
\begin{table}[ht]
    \centering
    \caption{\textbf{Comparison of dosimetric endpoints across planning strategies for patient P2 on \nth{1} rpCT.} All dose values ($D_{\text{0\%/0.03\si{cc}/mean}}$) are in~\si{\rbe}.}
    \label{tab:P2_metrics}
    \small
    \resizebox{0.9\textwidth}{!}{%
        \begin{tabular}{lllc cc cc cc}
            \toprule
            \textbf{Delineated} & \multicolumn{2}{c}{\textbf{Dosimetric Endpoint}} & \multicolumn{2}{c}{\textbf{Manual}} & \multicolumn{2}{c}{\textbf{DQN}} & \multicolumn{2}{c}{\textbf{PPO}} \\
            \cmidrule(lr){2-3} \cmidrule(lr){4-5} \cmidrule(lr){6-7} \cmidrule(lr){8-9}
            \textbf{Structure} & \textbf{Metric} & \textbf{Score} & \textbf{Value} & \textbf{Score} & \textbf{Value} & \textbf{Score} & \textbf{Value} & \textbf{Score} \\
            \midrule
            \multirow{2}{*}{CTV1} 
            & $V_{d_{Rx,\text{CTV}_1}} \geq 98\%$ & 15 & 97.99 & 14.99 & 98.01 & 15.00 & 97.99 & 14.99 \\
            & $D_{max} \leq 77$ & 5 & 82.30 & 3.24 & 83.03 & 3.14 & 83.61 & 3.06 \\
            \midrule
            CTV2 & $V_{d_{Rx,\text{CTV}_2}} \geq 98\%$ & 15 & 97.94 & 14.96 & 98.27 & 15.00 & 98.04 & 15.00 \\
            CTV3 & $V_{d_{Rx,\text{CTV}_3}} \geq 98\%$ & 15 & 97.76 & 14.84 & 98.27 & 15.00 & 98.18 & 15.00 \\
            \midrule
            BRS & $D_{\text{0.03cc}} \leq 30$ & 21 & 19.59 & 21.00 & 19.58 & 21.00 & 20.04 & 21.00 \\
            SC & $D_{\text{0.03cc}} \leq 30$ & 21 & 22.63 & 21.00 & 22.70 & 21.00 & 22.85 & 21.00 \\
            MAN & $V_{70\si{\rbe}} \leq 10\%$ & 5 & 1.29 & 5.00 & 1.07 & 5.00 & 1.51 & 5.00 \\
            LAR & $D_{\text{mean}} \leq 40$ & 7 & 36.51 & 7.00 & 36.48 & 7.00 & 36.76 & 7.00 \\
            PHY & $D_{\text{mean}} \leq 50$ & 7 & 49.13 & 7.00 & 48.95 & 7.00 & 49.41 & 7.00 \\
            PARL & $D_{\text{mean}} \leq 26$ & 7 & 26.00 & 7.00 & 26.00 & 7.00 & 22.74 & 7.00 \\
            PARR & $D_{\text{mean}} \leq 26$ & 7 & 28.41 & 2.78 & 25.15 & 7.00 & 25.81 & 7.00 \\
            COCHL & $D_{\text{mean}} \leq 35$ & 5 & 39.34 & 3.27 & 39.82 & 3.07 & 33.55 & 5.00 \\
            COCHR & $D_{\text{mean}} \leq 35$ & 5 & 8.44 & 5.00 & 8.47 & 5.00 & 8.56 & 5.00 \\
            SMGL & $D_{\text{mean}} \leq 35$ & 5 & 72.57 & 0 & 72.11 & 0 & 72.68 & 0 \\
            SMGR & $D_{\text{mean}} \leq 35$ & 5 & 72.83 & 0 & 72.51 & 0 & 73.06 & 0 \\
            ESO & $D_{\text{mean}} \leq 40$ & 5 & 6.23 & 5.00 & 6.26 & 5.00 & 6.24 & 5.00 \\
            \midrule
            \multicolumn{2}{l}{\textbf{Cumulative Total}} & \textbf{150} & \textbf{--} & 132.09 & \textbf{--} & 136.20 & \textbf{--} & \textbf{138.05} \\ 
            \bottomrule
        \end{tabular}
    }
\end{table}

\begin{figure}[ht]
    \centering 
    \includegraphics[width=0.9\linewidth]{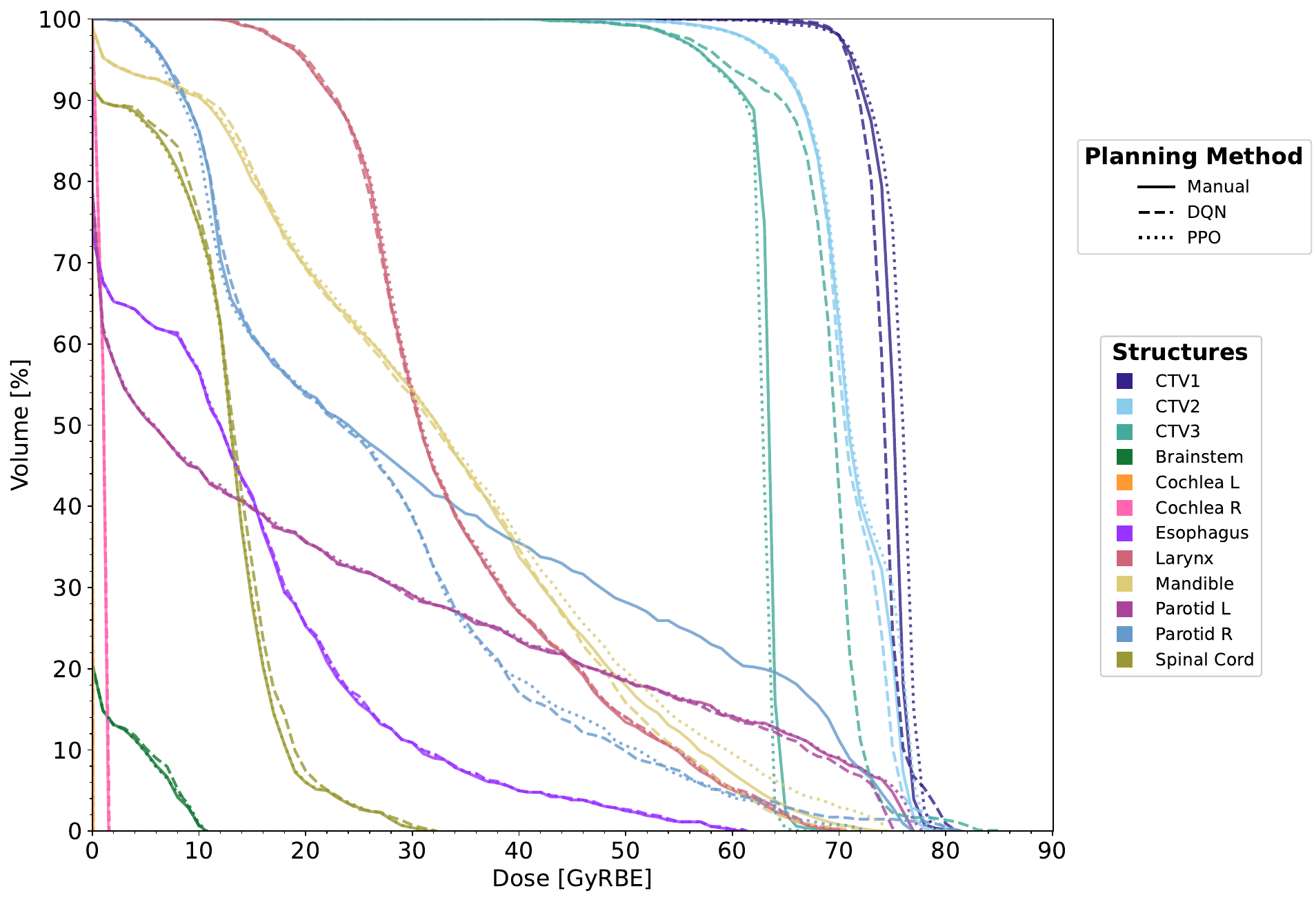} 
    \caption{\textbf{DVH comparison for patient P3's \nth{1} rpCT:} Manual replans (solid lines), patient-specific DQN (dashed lines), and patient-specific PPO (dotted lines).}
    \label{fig:P3_dvh} 
\end{figure}

\begin{table}[ht]
    \centering
    \caption{\textbf{Comparison of dosimetric endpoints across planning strategies for patient P3 on \nth{1} rpCT.} All dose values ($D_{\text{0\%/0.03\si{cc}/mean}}$) are in~\si{\rbe}.}
    \label{tab:P3_metrics}
    \small
    \resizebox{0.9\textwidth}{!}{%
        \begin{tabular}{lllc cc cc cc}
            \toprule
            \textbf{Delineated} & \multicolumn{2}{c}{\textbf{Dosimetric Endpoint}} & \multicolumn{2}{c}{\textbf{Manual}} & \multicolumn{2}{c}{\textbf{DQN}} & \multicolumn{2}{c}{\textbf{PPO}} \\
            \cmidrule(lr){2-3} \cmidrule(lr){4-5} \cmidrule(lr){6-7} \cmidrule(lr){8-9}
            \textbf{Structure} & \textbf{Metric} & \textbf{Score} & \textbf{Value} & \textbf{Score} & \textbf{Value} & \textbf{Score} & \textbf{Value} & \textbf{Score} \\
            \midrule
            \multirow{2}{*}{CTV1} 
            & $V_{d_{Rx,\text{CTV}_1}} \geq 98\%$ & 15 & 97.99 & 14.99 & 98.00 & 15.00 & 97.99 & 14.99 \\
            & $D_{max} \leq 77$ & 5 & 81.52 & 3.35 & 81.00 & 3.43 & 79.68 & 3.62 \\
            \midrule
            CTV2 & $V_{d_{Rx,\text{CTV}_2}} \geq 98\%$ & 15 & 98.39 & 15.00 & 98.39 & 15.00 & 98.29 & 15.00 \\
            CTV3 & $V_{d_{Rx,\text{CTV}_3}} \geq 98\%$ & 15 & 98.18 & 15.00 & 98.33 & 15.00 & 98.02 & 15.00 \\
            \midrule
            BRS & $D_{\text{0.03cc}} \leq 30$ & 21 & 10.69 & 21.00 & 10.55 & 21.00 & 10.77 & 21.00 \\
            SC & $D_{\text{0.03cc}} \leq 30$ & 21 & 31.81 & 20.09 & 32.35 & 19.82 & 32.17 & 19.91 \\
            MAN & $V_{70\si{\rbe}} \leq 10\%$ & 5 & 0.13 & 5.00 & 0.00 & 5.00 & 1.28 & 5.00 \\
            LAR & $D_{\text{mean}} \leq 40$ & 7 & 34.76 & 7.00 & 34.84 & 7.00 & 34.91 & 7.00 \\
            PHY & $D_{\text{mean}} \leq 50$ & 7 & 0.00 & 7.00 & 0.00 & 7.00 & 0.00 & 7.00 \\
            PARL & $D_{\text{mean}} \leq 26$ & 7 & 20.39 & 7.00 & 20.19 & 7.00 & 20.52 & 7.00 \\
            PARR & $D_{\text{mean}} \leq 26$ & 7 & 32.68 & 0.00 & 25.97 & 7.00 & 25.71 & 7.00 \\
            COCHL & $D_{\text{mean}} \leq 35$ & 5 & 0.02 & 5.00 & 0.02 & 5.00 & 0.02 & 5.00 \\
            COCHR & $D_{\text{mean}} \leq 35$ & 5 & 0.84 & 5.00 & 0.88 & 5.00 & 0.84 & 5.00 \\
            SMGL & $D_{\text{mean}} \leq 35$ & 5 & 0.00 & 5.00 & 0.00 & 5.00 & 0.00 & 5.00 \\
            SMGR & $D_{\text{mean}} \leq 35$ & 5 & 0.00 & 5.00 & 0.00 & 5.00 & 0.00 & 5.00 \\
            ESO & $D_{\text{mean}} \leq 40$ & 5 & 13.26 & 5.00 & 13.35 & 5.00 & 13.31 & 5.00 \\
            \midrule
            \multicolumn{2}{l}{\textbf{Cumulative Total}} & \textbf{150} & \textbf{--} & 140.44 & \textbf{--} & 147.25 & \textbf{--} & \textbf{147.52} \\ 
            \bottomrule
        \end{tabular}
    }
\end{table}

\begin{figure}[ht]
    \centering 
    \includegraphics[width=0.9\linewidth]{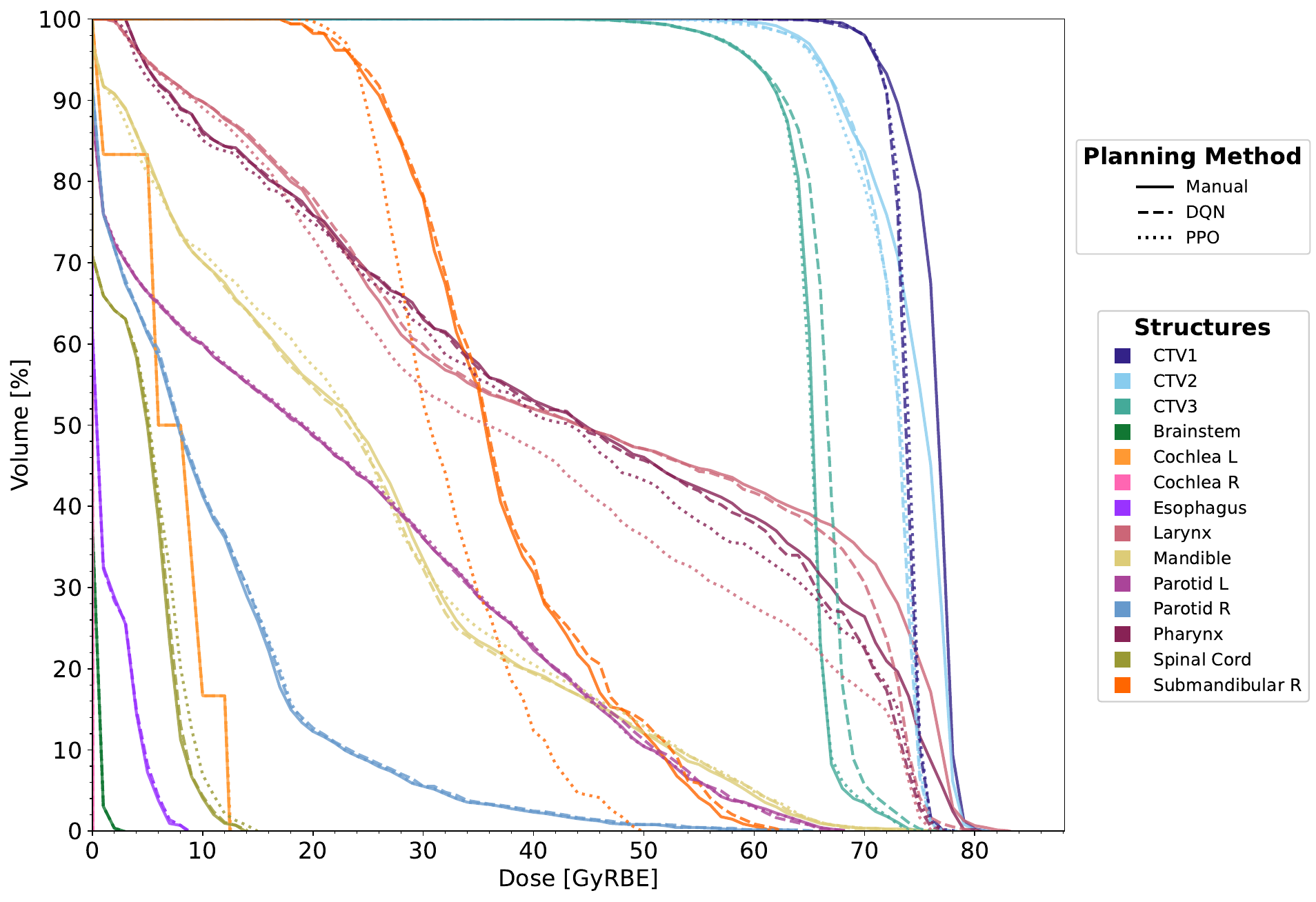} 
    \caption{\textbf{DVH comparison for patient P4's \nth{1} rpCT:} Manual replans (solid lines), patient-specific DQN (dashed lines), and patient-specific PPO (dotted lines).}
    \label{fig:P4_dvh} 
\end{figure}

\begin{table}[ht]
    \centering
    \caption{\textbf{Comparison of dosimetric endpoints across planning strategies for patient P4 on \nth{1} rpCT.} All dose values ($D_{\text{0\%/0.03\si{cc}/mean}}$) are in~\si{\rbe}.}
    \label{tab:P4_metrics}
    \small
    \resizebox{0.9\textwidth}{!}{%
        \begin{tabular}{lllc cc cc cc}
            \toprule
            \textbf{Delineated} & \multicolumn{2}{c}{\textbf{Dosimetric Endpoint}} & \multicolumn{2}{c}{\textbf{Manual}} & \multicolumn{2}{c}{\textbf{DQN}} & \multicolumn{2}{c}{\textbf{PPO}} \\
            \cmidrule(lr){2-3} \cmidrule(lr){4-5} \cmidrule(lr){6-7} \cmidrule(lr){8-9}
            \textbf{Structure} & \textbf{Metric} & \textbf{Score} & \textbf{Value} & \textbf{Score} & \textbf{Value} & \textbf{Score} & \textbf{Value} & \textbf{Score} \\
            \midrule
            \multirow{2}{*}{CTV1} 
            & $V_{d_{Rx,\text{CTV}_1}} \geq 98\%$ & 15 & 98.01 & 15.00 & 97.97 & 14.98 & 98.01 & 15.00 \\
            & $D_{max} \leq 77$ & 5 & 81.23 & 3.40 & 77.82 & 3.88 & 78.58 & 3.77 \\
            \midrule
            CTV2 & $V_{d_{Rx,\text{CTV}_2}} \geq 98\%$ & 15 & 98.55 & 15.00 & 98.14 & 15.00 & 98.19 & 15.00 \\
            CTV3 & $V_{d_{Rx,\text{CTV}_3}} \geq 98\%$ & 15 & 98.02 & 15.00 & 97.94 & 14.96 & 98.03 & 15.00 \\
            \midrule
            BRS & $D_{\text{0.03cc}} \leq 30$ & 21 & 2.87 & 21.00 & 2.82 & 21.00 & 2.85 & 21.00 \\
            SC & $D_{\text{0.03cc}} \leq 30$ & 21 & 13.85 & 21.00 & 14.04 & 21.00 & 14.98 & 21.00 \\
            MAN & $V_{70\si{\rbe}} \leq 10\%$ & 5 & 0.22 & 5.00 & 0.09 & 5.00 & 0.09 & 5.00 \\
            LAR & $D_{\text{mean}} \leq 40$ & 7 & 45.29 & 3.30 & 44.51 & 5.20 & 39.29 & 7.00 \\
            PHY & $D_{\text{mean}} \leq 50$ & 7 & 44.00 & 7.00 & 43.17 & 7.00 & 42.13 & 7.00 \\
            PARL & $D_{\text{mean}} \leq 26$ & 7 & 21.76 & 7.00 & 21.79 & 7.00 & 21.86 & 7.00 \\
            PARR & $D_{\text{mean}} \leq 26$ & 7 & 10.08 & 7.00 & 10.28 & 7.00 & 10.19 & 7.00 \\
            COCHL & $D_{\text{mean}} \leq 35$ & 5 & 6.92 & 5.00 & 6.93 & 5.00 & 6.97 & 5.00 \\
            COCHR & $D_{\text{mean}} \leq 35$ & 5 & 0.02 & 5.00 & 0.02 & 5.00 & 0.02 & 5.00 \\
            SMGL & $D_{\text{mean}} \leq 35$ & 5 & 0.00 & 5.00 & 0.00 & 5.00 & 0.00 & 5.00 \\
            SMGR & $D_{\text{mean}} \leq 35$ & 5 & 36.87 & 4.25 & 37.30 & 4.08 & 31.98 & 5.00 \\
            ESO & $D_{\text{mean}} \leq 40$ & 5 & 1.36 & 5.00 & 1.38 & 5.00 & 1.38 & 5.00 \\
            \midrule
            \multicolumn{2}{l}{\textbf{Cumulative Total}} & \textbf{150} & \textbf{--} & 143.94 & \textbf{--} & 146.10 & \textbf{--} & \textbf{148.77} \\ 
            \bottomrule
        \end{tabular}
    }
\end{table}

\begin{figure}[ht]
    \centering 
    \includegraphics[width=0.9\linewidth]{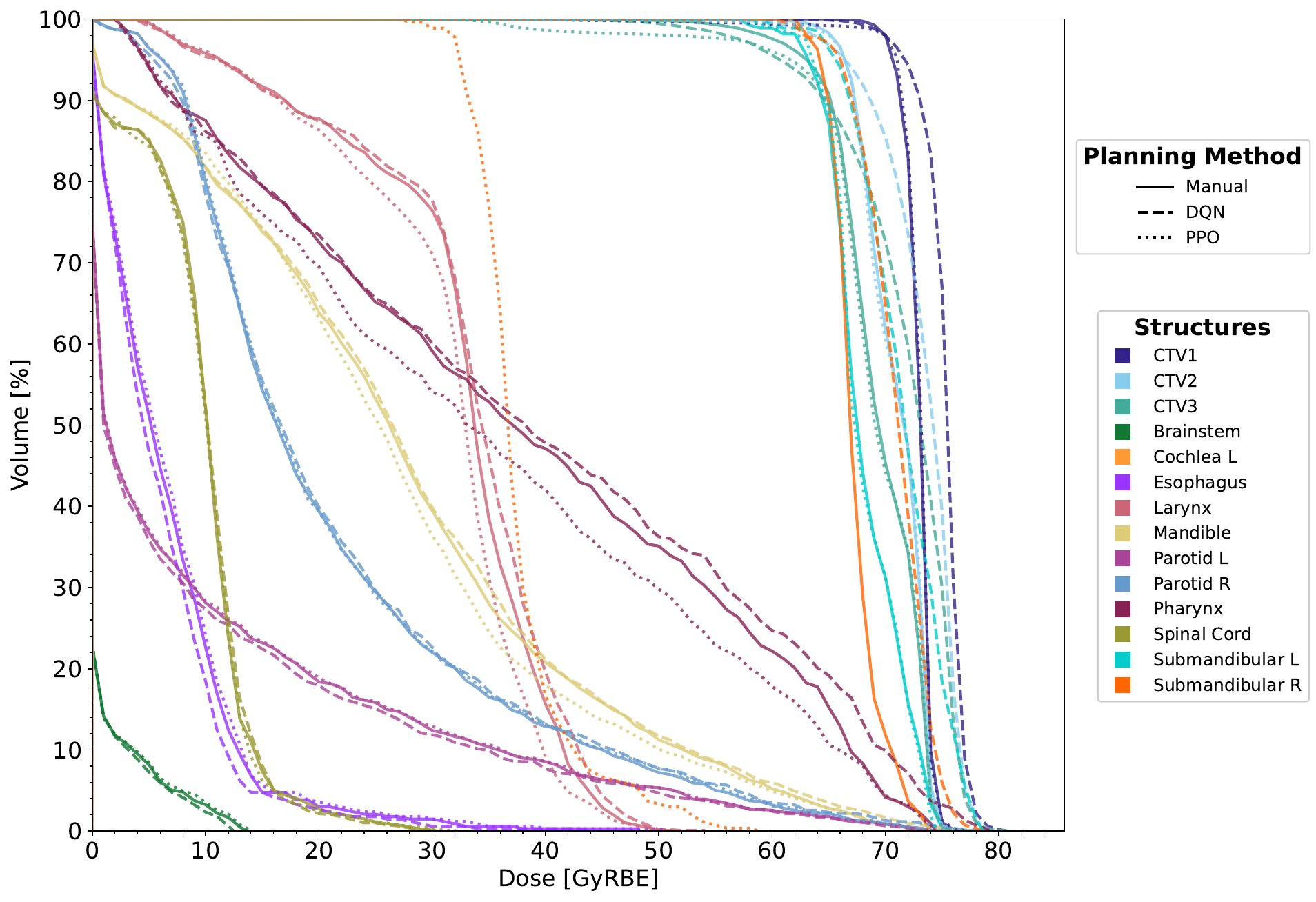} 
    \caption{\textbf{DVH comparison for patient P5's \nth{1} rpCT:} Manual replans (solid lines), patient-specific DQN (dashed lines), and patient-specific PPO (dotted lines).}
    \label{fig:P5_dvh} 
\end{figure}

\begin{table}[ht]
    \centering
    \caption{\textbf{Comparison of dosimetric endpoints across planning strategies for patient P5 on \nth{1} rpCT.} All dose values ($D_{\text{0\%/0.03\si{cc}/mean}}$) are in~\si{\rbe}.}
    \label{tab:P5_metrics}
    \small
    \resizebox{0.9\textwidth}{!}{%
        \begin{tabular}{lllc cc cc cc}
            \toprule
            \textbf{Delineated} & \multicolumn{2}{c}{\textbf{Dosimetric Endpoint}} & \multicolumn{2}{c}{\textbf{Manual}} & \multicolumn{2}{c}{\textbf{DQN}} & \multicolumn{2}{c}{\textbf{PPO}} \\
            \cmidrule(lr){2-3} \cmidrule(lr){4-5} \cmidrule(lr){6-7} \cmidrule(lr){8-9}
            \textbf{Structure} & \textbf{Metric} & \textbf{Score} & \textbf{Value} & \textbf{Score} & \textbf{Value} & \textbf{Score} & \textbf{Value} & \textbf{Score} \\
            \midrule
            \multirow{2}{*}{CTV1} 
            & $V_{d_{Rx,\text{CTV}_1}} \geq 98\%$ & 15 & 97.99 & 14.99 & 97.99 & 14.99 & 98.00 & 15.00 \\
            & $D_{max} \leq 77$ & 5 & 77.26 & 3.96 & 80.82 & 3.45 & 77.82 & 3.88 \\
            \midrule
            CTV2 & $V_{d_{Rx,\text{CTV}_2}} \geq 98\%$ & 15 & 99.51 & 15.00 & 98.50 & 15.00 & 99.19 & 15.00 \\
            CTV3 & $V_{d_{Rx,\text{CTV}_3}} \geq 98\%$ & 15 & 99.12 & 15.00 & 97.98 & 14.99 & 97.53 & 14.69 \\
            \midrule
            BRS & $D_{\text{0.03cc}} \leq 30$ & 21 & 13.69 & 21.00 & 12.48 & 21.00 & 14.07 & 21.00 \\
            SC & $D_{\text{0.03cc}} \leq 30$ & 21 & 30.58 & 20.71 & 30.23 & 20.88 & 32.06 & 19.97 \\
            MAN & $V_{70\si{\rbe}} \leq 10\%$ & 5 & 0.08 & 5.00 & 0.67 & 5.00 & 0.20 & 5.00 \\
            LAR & $D_{\text{mean}} \leq 40$ & 7 & 32.10 & 7.00 & 32.74 & 7.00 & 30.86 & 7.00 \\
            PHY & $D_{\text{mean}} \leq 50$ & 7 & 37.73 & 7.00 & 38.98 & 7.00 & 35.12 & 7.00 \\
            PARL & $D_{\text{mean}} \leq 26$ & 7 & 10.04 & 7.00 & 9.63 & 7.00 & 10.12 & 7.00 \\
            PARR & $D_{\text{mean}} \leq 26$ & 7 & 21.56 & 7.00 & 21.81 & 7.00 & 21.69 & 7.00 \\
            COCHL & $D_{\text{mean}} \leq 35$ & 5 & 0.00 & 5.00 & 0.00 & 5.00 & 0.00 & 5.00 \\
            COCHR & $D_{\text{mean}} \leq 35$ & 5 & 0.00 & 5.00 & 0.00 & 5.00 & 0.00 & 5.00 \\
            SMGL & $D_{\text{mean}} \leq 35$ & 5 & 68.14 & 0 & 71.57 & 0 & 68.15 & 0 \\
            SMGR & $D_{\text{mean}} \leq 35$ & 5 & 67.28 & 0 & 70.89 & 0 & 37.60 & 0 \\
            ESO & $D_{\text{mean}} \leq 40$ & 5 & 6.41 & 5.00 & 5.85 & 5.00 & 6.69 & 5.00 \\
            \midrule
            \multicolumn{2}{l}{\textbf{Cumulative Total}} & \textbf{150} & \textbf{--} & 138.67 & \textbf{--} & 138.32 & \textbf{--} & \textbf{141.50} \\ 
            \bottomrule
        \end{tabular}
    }
\end{table}

\begin{figure}[ht]
    \centering 
    \includegraphics[width=0.9\linewidth]{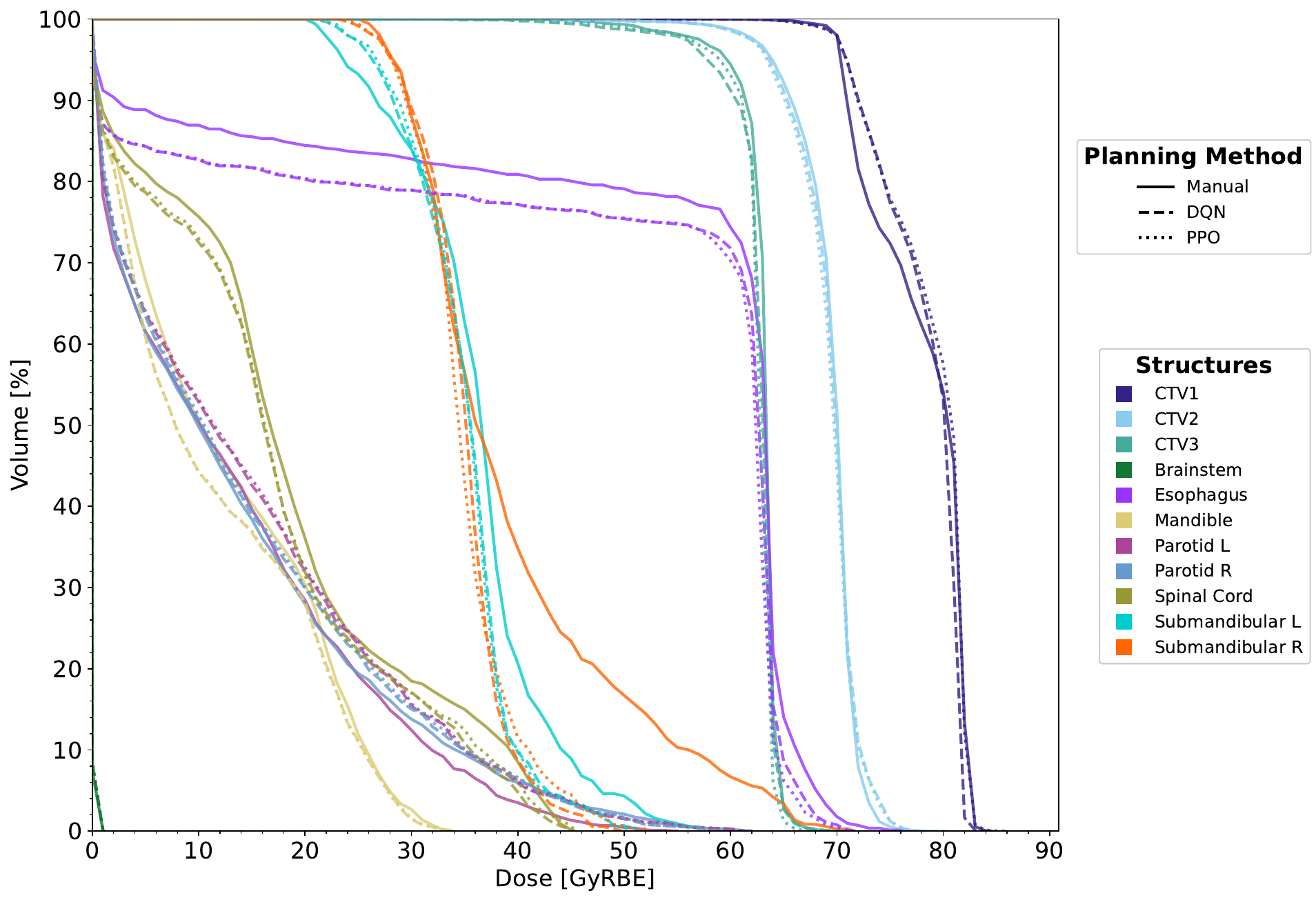} 
    \caption{\textbf{DVH comparison for patient P6's \nth{1} rpCT:} Manual replans (solid lines), patient-specific DQN (dashed lines), and patient-specific PPO (dotted lines).}
    \label{fig:P6_dvh} 
\end{figure}

\begin{table}[ht]
    \centering
    \caption{\textbf{Comparison of dosimetric endpoints across planning strategies for patient P6 on \nth{1} rpCT.} All dose values ($D_{\text{0\%/0.03\si{cc}/mean}}$) are in~\si{\rbe}.}
    \label{tab:P6_metrics}
    \small
    \resizebox{0.9\textwidth}{!}{%
        \begin{tabular}{lllc cc cc cc}
            \toprule
            \textbf{Delineated} & \multicolumn{2}{c}{\textbf{Dosimetric Endpoint}} & \multicolumn{2}{c}{\textbf{Manual}} & \multicolumn{2}{c}{\textbf{DQN}} & \multicolumn{2}{c}{\textbf{PPO}} \\
            \cmidrule(lr){2-3} \cmidrule(lr){4-5} \cmidrule(lr){6-7} \cmidrule(lr){8-9}
            \textbf{Structure} & \textbf{Metric} & \textbf{Score} & \textbf{Value} & \textbf{Score} & \textbf{Value} & \textbf{Score} & \textbf{Value} & \textbf{Score} \\
            \midrule
            \multirow{2}{*}{CTV1} 
            & $V_{d_{Rx,\text{CTV}_1}} \geq 98\%$ & 15 & 98.00 & 15.00 & 98.01 & 15.00 & 98.00 & 15.00 \\
            & $D_{max} \leq 77$ & 5 & 84.54 & 2.92 & 84.74 & 2.89 & 85.85 & 2.74 \\
            \midrule
            CTV2 & $V_{d_{Rx,\text{CTV}_2}} \geq 98\%$ & 15 & 98.69 & 15.00 & 98.41 & 15.00 & 98.51 & 15.00 \\
            CTV3 & $V_{d_{Rx,\text{CTV}_3}} \geq 98\%$ & 15 & 98.46 & 15.00 & 98.15 & 15.00 & 98.39 & 15.00 \\
            \midrule
            BRS & $D_{\text{0.03cc}} \leq 30$ & 21 & 1.11 & 21.00 & 1.03 & 21.00 & 1.23 & 21.00 \\
            SC & $D_{\text{0.03cc}} \leq 30$ & 21 & 45.13 & 13.44 & 45.38 & 13.31 & 45.43 & 13.29 \\
            MAN & $V_{70\si{\rbe}} \leq 10\%$ & 5 & 0.00 & 5.00 & 0.00 & 5.00 & 0.00 & 5.00 \\
            LAR & $D_{\text{mean}} \leq 40$ & 7 & 0.00 & 7.00 & 0.00 & 7.00 & 0.00 & 7.00 \\
            PHY & $D_{\text{mean}} \leq 50$ & 7 & 0.00 & 7.00 & 0.00 & 7.00 & 0.00 & 7.00 \\
            PARL & $D_{\text{mean}} \leq 26$ & 7 & 13.20 & 7.00 & 14.66 & 7.00 & 14.80 & 7.00 \\
            PARR & $D_{\text{mean}} \leq 26$ & 7 & 13.67 & 7.00 & 14.13 & 7.00 & 14.31 & 7.00 \\
            COCHL & $D_{\text{mean}} \leq 35$ & 5 & 0.00 & 5.00 & 0.00 & 5.00 & 0.00 & 5.00 \\
            COCHR & $D_{\text{mean}} \leq 35$ & 5 & 0.00 & 5.00 & 0.00 & 5.00 & 0.00 & 5.00 \\
            SMGL & $D_{\text{mean}} \leq 35$ & 5 & 36.19 & 4.53 & 34.94 & 5.00 & 34.97 & 5.00 \\
            SMGR & $D_{\text{mean}} \leq 35$ & 5 & 39.59 & 3.16 & 35.01 & 5.00 & 34.96 & 5.00 \\
            ESO & $D_{\text{mean}} \leq 40$ & 5 & 52.75 & 0.00 & 49.92 & 0.04 & 49.69 & 0.15 \\
            \midrule
            \multicolumn{2}{l}{\textbf{Cumulative Total}} & \textbf{150} & \textbf{--} & 133.03 & \textbf{--} & \textbf{135.24} & \textbf{--} & 135.18 \\ 
            \bottomrule
        \end{tabular}
    }
\end{table}

\begin{figure}[ht]
    \centering 
    \includegraphics[width=0.9\linewidth]{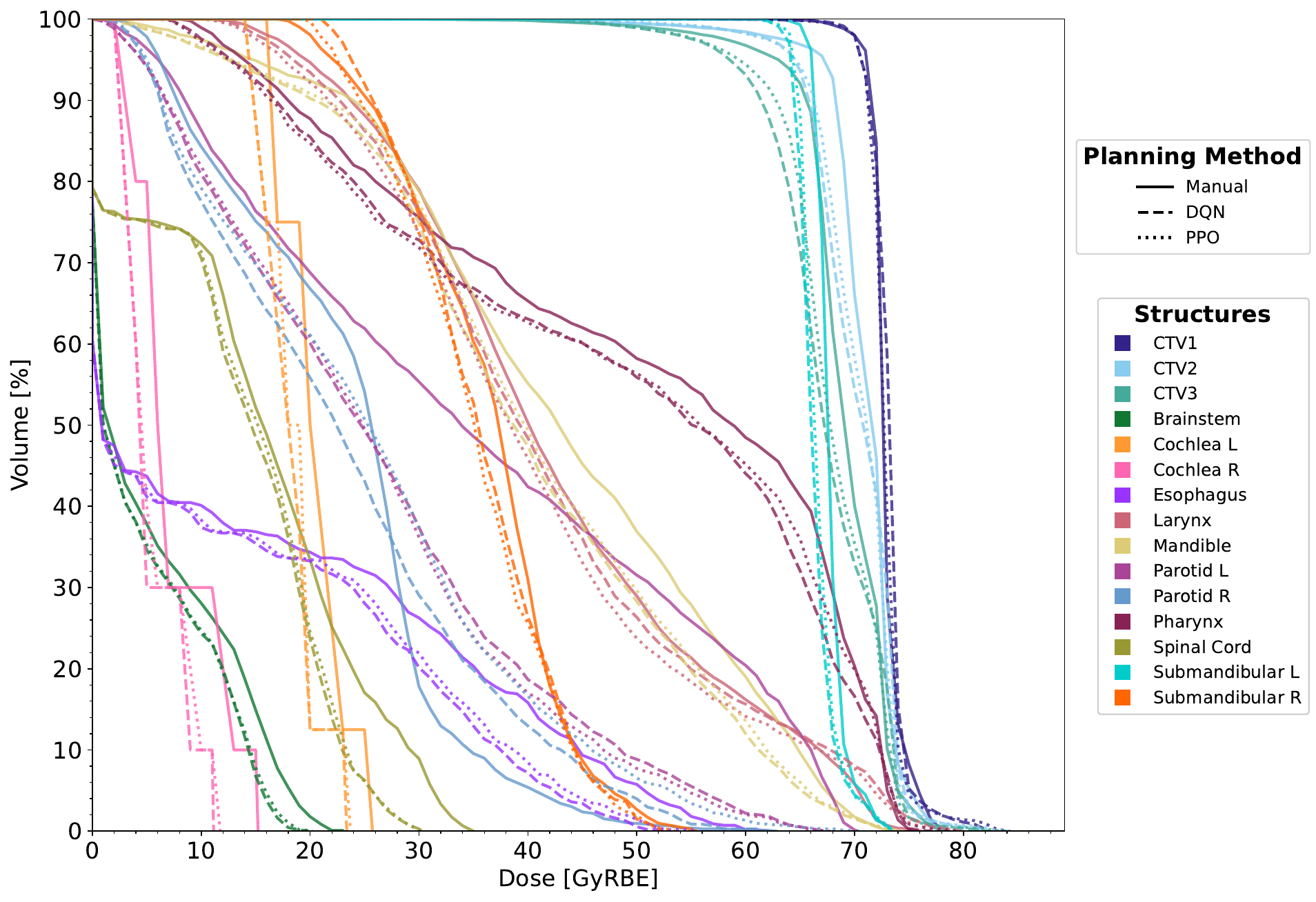} 
    \caption{\textbf{DVH comparison for patient P7's \nth{1} rpCT:} Manual replans (solid lines), patient-specific DQN (dashed lines), and patient-specific PPO (dotted lines).}
    \label{fig:P7_dvh} 
\end{figure}

\begin{table}[ht]
    \centering
    \caption{\textbf{Comparison of dosimetric endpoints across planning strategies for patient P7 on \nth{1} rpCT.} All dose values ($D_{\text{0\%/0.03\si{cc}/mean}}$) are in~\si{\rbe}.}
    \label{tab:P7_metrics}
    \small
    \resizebox{0.9\textwidth}{!}{%
        \begin{tabular}{lllc cc cc cc}
            \toprule
            \textbf{Delineated} & \multicolumn{2}{c}{\textbf{Dosimetric Endpoint}} & \multicolumn{2}{c}{\textbf{Manual}} & \multicolumn{2}{c}{\textbf{DQN}} & \multicolumn{2}{c}{\textbf{PPO}} \\
            \cmidrule(lr){2-3} \cmidrule(lr){4-5} \cmidrule(lr){6-7} \cmidrule(lr){8-9}
            \textbf{Structure} & \textbf{Metric} & \textbf{Score} & \textbf{Value} & \textbf{Score} & \textbf{Value} & \textbf{Score} & \textbf{Value} & \textbf{Score} \\
            \midrule
            \multirow{2}{*}{CTV1} 
            & $V_{d_{Rx,\text{CTV}_1}} \geq 98\%$ & 15 & 98.01 & 15.00 & 97.99 & 14.99 & 97.99 & 14.99 \\
            & $D_{max} \leq 77$ & 5 & 78.64 & 3.77 & 82.44 & 3.22 & 84.30 & 2.96 \\
            \midrule
            CTV2 & $V_{d_{Rx,\text{CTV}_2}} \geq 98\%$ & 15 & 97.63 & 14.76 & 96.98 & 14.32 & 96.86 & 14.24 \\
            CTV3 & $V_{d_{Rx,\text{CTV}_3}} \geq 98\%$ & 15 & 98.08 & 15.00 & 97.16 & 14.44 & 97.30 & 14.53 \\
            \midrule
            BRS & $D_{\text{0.03cc}} \leq 30$ & 21 & 23.07 & 21.00 & 19.85 & 21.00 & 20.23 & 21.00 \\
            SC & $D_{\text{0.03cc}} \leq 30$ & 21 & 34.99 & 18.50 & 30.29 & 20.85 & 30.27 & 20.87 \\
            MAN & $V_{70\si{\rbe}} \leq 10\%$ & 5 & 0.03 & 5.00 & 0.07 & 5.00 & 0.21 & 5.00 \\
            LAR & $D_{\text{mean}} \leq 40$ & 7 & 42.36 & 6.05 & 41.64 & 6.34 & 40.61 & 6.76 \\
            PHY & $D_{\text{mean}} \leq 50$ & 7 & 50.47 & 6.81 & 48.31 & 7.00 & 48.50 & 7.00 \\
            PARL & $D_{\text{mean}} \leq 26$ & 7 & 35.68 & 0.00 & 25.86 & 7.00 & 25.61 & 7.00 \\
            PARR & $D_{\text{mean}} \leq 26$ & 7 & 23.26 & 7.00 & 23.10 & 7.00 & 25.32 & 7.00 \\
            COCHL & $D_{\text{mean}} \leq 35$ & 5 & 20.38 & 5.00 & 18.11 & 5.00 & 18.43 & 5.00 \\
            COCHR & $D_{\text{mean}} \leq 35$ & 5 & 7.54 & 5.00 & 5.31 & 5.00 & 5.60 & 5.00 \\
            SMGL & $D_{\text{mean}} \leq 35$ & 5 & 67.71 & 5.00 & 66.30 & 0.00 & 66.60 & 0.00 \\
            SMGR & $D_{\text{mean}} \leq 35$ & 5 & 35.91 & 4.63 & 35.35 & 4.86 & 34.84 & 5.00 \\
            ESO & $D_{\text{mean}} \leq 40$ & 5 & 14.48 & 5.00 & 12.24 & 5.00 & 12.65 & 5.00 \\
            \midrule
            \multicolumn{2}{l}{\textbf{Cumulative Total}} & \textbf{150} & \textbf{--} & 132.53 & \textbf{--} & 141.03 & \textbf{--} & \textbf{141.35} \\ 
            \bottomrule
        \end{tabular}
    }
\end{table}

\begin{figure}[ht]
    \centering 
    \includegraphics[width=0.9\linewidth]{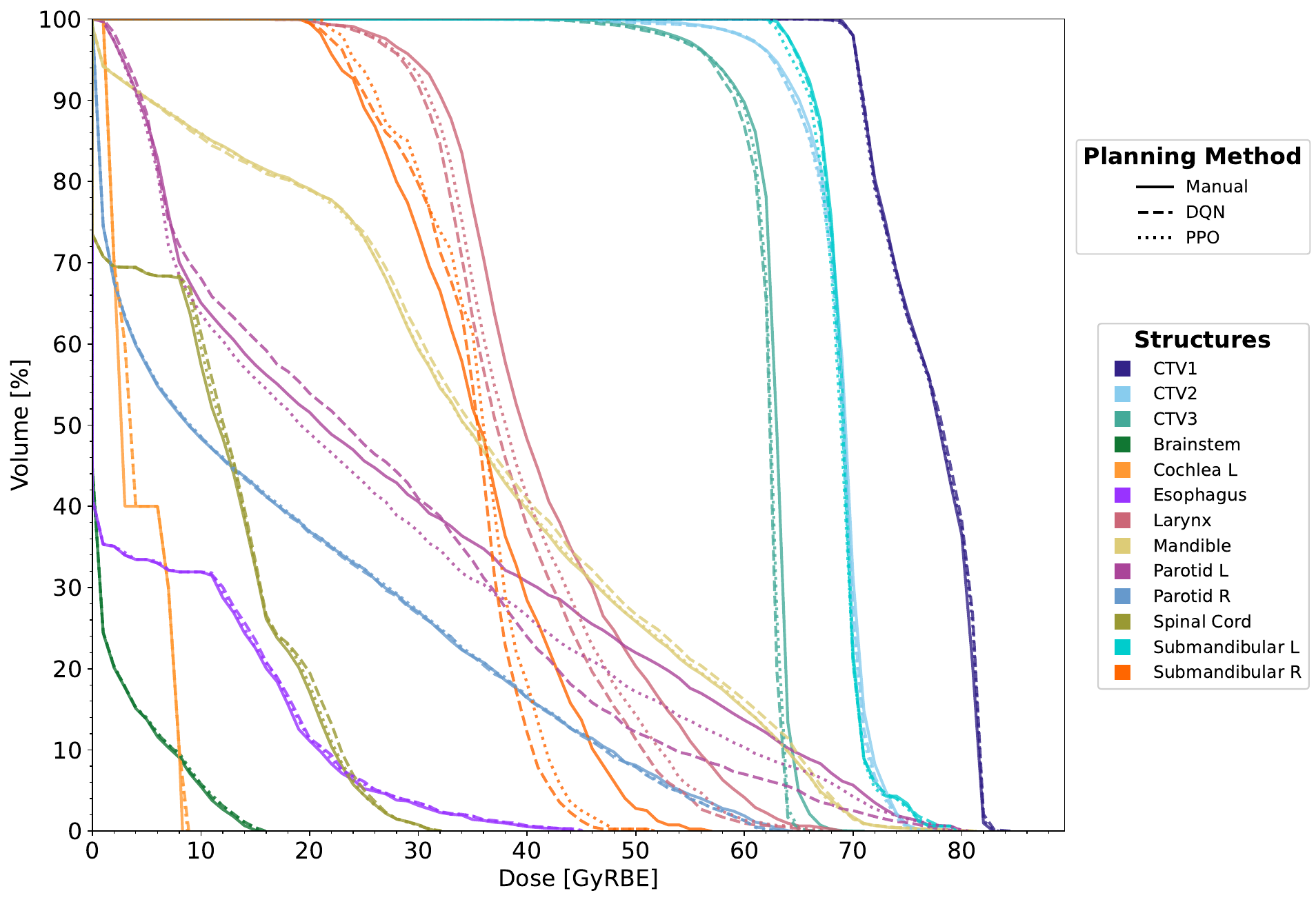} 
    \caption{\textbf{DVH comparison for patient P8's \nth{1} rpCT:} Manual replans (solid lines), patient-specific DQN (dashed lines), and patient-specific PPO (dotted lines).}
    \label{fig:P8_dvh} 
\end{figure}

\begin{table}[ht]
    \centering
    \caption{\textbf{Comparison of dosimetric endpoints across planning strategies for patient P8 on \nth{1} rpCT.} All dose values ($D_{\text{0\%/0.03\si{cc}/mean}}$) are in~\si{\rbe}.}
    \label{tab:P8_metrics}
    \small
    \resizebox{0.9\textwidth}{!}{%
        \begin{tabular}{lllc cc cc cc}
            \toprule
            \textbf{Delineated} & \multicolumn{2}{c}{\textbf{Dosimetric Endpoint}} & \multicolumn{2}{c}{\textbf{Manual}} & \multicolumn{2}{c}{\textbf{DQN}} & \multicolumn{2}{c}{\textbf{PPO}} \\
            \cmidrule(lr){2-3} \cmidrule(lr){4-5} \cmidrule(lr){6-7} \cmidrule(lr){8-9}
            \textbf{Structure} & \textbf{Metric} & \textbf{Score} & \textbf{Value} & \textbf{Score} & \textbf{Value} & \textbf{Score} & \textbf{Value} & \textbf{Score} \\
            \midrule
            \multirow{2}{*}{CTV1} 
            & $V_{d_{Rx,\text{CTV}_1}} \geq 98\%$ & 15 & 97.99 & 14.99 & 97.99 & 14.99 & 98.02 & 15.00 \\
            & $D_{max} \leq 77$ & 5 & 82.89 & 3.16 & 84.47 & 2.93 & 83.44 & 3.08 \\
            \midrule
            CTV2 & $V_{d_{Rx,\text{CTV}_2}} \geq 98\%$ & 15 & 97.86 & 14.91 & 98.06 & 15.00 & 97.99 & 14.99 \\
            CTV3 & $V_{d_{Rx,\text{CTV}_3}} \geq 98\%$ & 15 & 97.77 & 14.85 & 97.51 & 14.68 & 97.51 & 14.68 \\
            \midrule
            BRS & $D_{\text{0.03cc}} \leq 30$ & 21 & 15.29 & 21.00 & 15.94 & 21.00 & 15.74 & 21.00 \\
            SC & $D_{\text{0.03cc}} \leq 30$ & 21 & 31.87 & 20.06 & 32.14 & 19.93 & 31.85 & 20.07 \\
            MAN & $V_{70\si{\rbe}} \leq 10\%$ & 5 & 0.46 & 5.00 & 0.52 & 5.00 & 0.46 & 5.00 \\
            LAR & $D_{\text{mean}} \leq 40$ & 7 & 41.84 & 6.26 & 38.99 & 7.00 & 39.90 & 7.00 \\
            PHY & $D_{\text{mean}} \leq 50$ & 7 & 0.00 & 7.00 & 0.00 & 7.00 & 0.00 & 7.00 \\
            PARL & $D_{\text{mean}} \leq 26$ & 7 & 28.00 & 5.00 & 25.98 & 7.00 & 25.86 & 7.00 \\
            PARR & $D_{\text{mean}} \leq 26$ & 7 & 17.32 & 7.00 & 17.20 & 7.00 & 17.31 & 7.00 \\
            COCHL & $D_{\text{mean}} \leq 35$ & 5 & 4.15 & 5.00 & 4.40 & 5.00 & 4.37 & 5.00 \\
            COCHR & $D_{\text{mean}} \leq 35$ & 5 & 0.00 & 5.00 & 0.00 & 5.00 & 0.00 & 5.00 \\
            SMGL & $D_{\text{mean}} \leq 35$ & 5 & 69.03 & 0.00 & 68.98 & 0.00 & 68.77 & 0.00 \\
            SMGR & $D_{\text{mean}} \leq 35$ & 5 & 35.51 & 4.80 & 34.32 & 5.00 & 35.11 & 4.96 \\
            ESO & $D_{\text{mean}} \leq 40$ & 5 & 6.37 & 5.00 & 6.53 & 5.00 & 6.46 & 5.00 \\
            \midrule
            \multicolumn{2}{l}{\textbf{Cumulative Total}} & \textbf{150} & \textbf{--} & 139.03 & \textbf{--} & 141.53 & \textbf{--} & \textbf{141.78} \\ 
            \bottomrule
        \end{tabular}
    }
\end{table}

\begin{table}[t]
    \centering
    \caption{\textbf{Summary of dosimetric performance across patients P6-P8 on \nth{1} rpCT:} Manual (M), patient-specific DQN (Q), and patient-specific PPO (P). All dose metrics ($D_{\text{0\%/0.03\si{cc}/mean}}$) in~\si{\rbe}. Bold values indicate superior dosimetry outcomes. Results for patients P1-P5 are summarized in Table~\ref{tab:full_results}.}  
    \label{tab:additional_dosimetric_results}
    \resizebox{0.95\textwidth}{!}{%
        \begin{tabular}{ll*{9}{c}}
            \toprule
            \multirow{2}{*}{\textbf{Structure}} & \multirow{2}{*}{\textbf{Metric}} & 
            \multicolumn{3}{c}{\textbf{P6}} & \multicolumn{3}{c}{\textbf{P7}} & 
            \multicolumn{3}{c}{\textbf{P8}} \\
            \cmidrule(lr){3-5} \cmidrule(lr){6-8} \cmidrule(lr){9-11}
            & & {M} & {Q} & {P} & {M} & {Q} & {P} & {M} & {Q} & {P} \\
            \midrule
            \multirow{2}{*}{CTV1} & $V_{d_{Rx,\text{CTV}_1}} \geq 98\%$ & 98.00 & \textbf{98.01} & 98.00 & 97.99 & 97.99 & \textbf{98.00} & \textbf{98.01} & 97.99 & 97.99 \\
            & $D_{0\%} \leq 77$ & \textbf{84.54} & 84.74 & 85.85 & \textbf{82.89} & 84.47 & 83.44 & \textbf{78.64} & 82.44 & 84.30 \\
            \midrule
            CTV2 & $V_{d_{Rx,\text{CTV}_2}} \geq 98\%$ & \textbf{98.69} & 98.41 & 98.51 & 97.86 & \textbf{98.06} & 97.99 & \textbf{97.63} & 96.98 & 96.86 \\
            CTV3 & $V_{d_{Rx,\text{CTV}_3}} \geq 98\%$ & \textbf{98.46} & 98.15 & 98.39 & \textbf{97.77} & 97.51 & 97.51 & \textbf{98.08} & 97.16 & 97.30 \\
            \midrule
            BRS & $D_{\text{0.03cc}} \leq 30$ & 1.11 & \textbf{1.03} & 1.23 & \textbf{15.29} & 15.94 & 15.74 & 23.07 & \textbf{19.85} & 20.23 \\
            SC  & $D_{\text{0.03cc}} \leq 30$ & \textbf{45.13} & 45.38 & 45.43 & 31.87 & 32.14 & \textbf{31.85} & 34.99 & 30.29 & \textbf{30.27} \\
            MAN & $V_{70\si{\rbe}} \leq 10\%$ & \textbf{0.00} & \textbf{0.00}  & \textbf{0.00}  & \textbf{0.46} & 0.52 & \textbf{0.46} & \textbf{0.03} & 0.07 & 0.21 \\
            LAR & $D_{\text{mean}} \leq 40$ & \textbf{0.00}  & \textbf{0.00}  & \textbf{0.00}  & 41.84 & \textbf{38.99} & 39.90 & 42.36 & 41.64 & \textbf{40.61} \\
            PHY & $D_{\text{mean}} \leq 50$ & \textbf{0.00}  & \textbf{0.00}  & \textbf{0.00}  & \textbf{0.00}  & \textbf{0.00}  & \textbf{0.00}  & 50.47 & \textbf{48.31} & 48.50 \\
            PARL & $D_{\text{mean}} \leq 26$ & \textbf{13.20} & 14.66 & 14.80 & 28.00 & 25.98 & \textbf{25.86} & 35.68 & 25.86 & \textbf{25.61} \\
            PARR & $D_{\text{mean}} \leq 26$ & \textbf{13.67} & 14.13 & 14.31 & 17.32 & \textbf{17.20} & 17.31 & 23.26 & \textbf{23.10} & 25.32 \\
            COCHL & $D_{\text{mean}} \leq 35$ & \textbf{0.00}  & \textbf{0.00}  & \textbf{0.00}  & \textbf{4.15} & 4.40 & 4.37 & 20.38 & \textbf{18.11} & 18.43 \\
            COCHR & $D_{\text{mean}} \leq 35$ & \textbf{0.00}  & \textbf{0.00}  & \textbf{0.00}  & \textbf{0.00}  & \textbf{0.00}  & \textbf{0.00}  & 7.54 & \textbf{5.31} & 5.60 \\
            SMGL & $D_{\text{mean}} \leq 35$ & 36.19 & \textbf{34.94} & 34.97 & 69.03 & 68.98 & \textbf{68.77} & 67.71 & \textbf{66.30} & 66.60 \\
            SMGR & $D_{\text{mean}} \leq 35$ & 39.59 & 35.01 & \textbf{34.96} & 35.51 & \textbf{34.32} & 35.11 & 35.91 & 35.35 & \textbf{34.84} \\
            ESO & $D_{\text{mean}} \leq 40$ & 52.57 & 49.92 & \textbf{49.69} & \textbf{6.37} & 6.53 & 6.46 & 14.48 & \textbf{12.24} & 12.65 \\
            \midrule
            \multicolumn{2}{l}{\textbf{Plan Score (max=150):}} & 133.03 & \textbf{135.24} & 135.18 & 132.53 & 141.03 & \textbf{141.35} & 139.03 & 141.53 & \textbf{141.78}\\
            \bottomrule
        \end{tabular}%
    }
    \begin{minipage}{0.9\textwidth}
        \footnotesize
        \vspace{0.5em}
        \textit{Note}: OAR abbreviations - BRS: Brainstem, SC: Spinal Cord, MAN: Mandible, LAR: Larynx, PHY: Pharynx, PARL/PARR: Left/Right Parotid, COCHL/COCHR: Left/Right Cochlea, SMLG/SMGR: Left/Right Submandibular Gland, ESO: Esophagus.
    \end{minipage}
\end{table}



\end{document}